\documentclass[floatfix,preprint,showpacs,preprintnumbers,superscriptaddress]{revtex4}
\usepackage{times} 
\usepackage{ifthen} 
\usepackage{graphicx}
\usepackage{amssymb,amsmath}
\graphicspath{{.}}
\newcommand{\mylab}[1]{\label{#1}}
\newcommand{\no}[1]{}
\begin{document}
\bibliographystyle{$HOME/literature/Bibdir/revtex}
\title{Self-propelled running droplets on solid substrates driven by chemical reactions}
\author{Karin John}
\email{john@pks.mpg.de}
\affiliation{Max-Planck-Institut
f\"ur Physik komplexer Systeme, N{\"o}thnitzer Str.\ 38, D-01187 Dresden, Germany}
\author{Markus B{\"a}r}
\email{markus.baer@ptb.de}
\affiliation{Physikalisch-Technische Bundesanstalt, Abbestr.\ 2--12, D-10587 Berlin, Germany}
\author{Uwe Thiele}
\email{thiele@pks.mpg.de, http:\\www.uwethiele.de}
\affiliation{Max-Planck-Institut
f\"ur Physik komplexer Systeme, N{\"o}thnitzer Str.\ 38, D-01187 Dresden, Germany}
\begin{abstract}
We study chemically driven running droplets on a partially wetting  
solid substrate by means of coupled evolution equations 
for the thickness profile of the droplets and the density profile 
of an adsorbate layer. 
Two models are introduced corresponding to two qualitatively different types of
experiments described in the literature.
In both cases an adsorption or desorption reaction underneath the droplets induces a
wettability gradient on the substrate and provides the driving 
force for droplet motion. The difference lies in the behavior of the
substrate behind the droplet. In case~I the substrate is irreversibly changed whereas
in case~II it recovers allowing for a periodic droplet
movement (as long as the overall system stays far away from equilibrium).
Both models allow for a non-saturated and a saturated regime of droplet movement 
depending on the ratio of the viscous and reactive time scales. 
In contrast to model~I, model~II allows for sitting drops at high reaction
rate and zero diffusion along the substrate. 
The transition from running to sitting drops in model~II occurs via a
super- or subcritical drift-pitchfork bifurcation and may be strongly hysteretic
implying a coexistence region of running and sitting drops.
\end{abstract}
\maketitle
%
%
\section{Introduction}
\mylab{intro}
%
The movement of droplets in external gradients fascinates scientists and
layman alike at least since Newton's description \cite{Newt1730hab2} of Hauksbee's
experiment with drops of orange oil that move between two non-parallel glass
plates towards the point of smallest plate distance \cite{Hauk1710}.
In another example a drop of liquid freely immersed in another liquid subject
to a temperature gradient will move towards the higher temperature region 
due to Marangoni forces caused by surface tension gradients \cite{Vela98}.
A drop sitting on a solid substrate also moves in a temperature \cite{Broc89}
or wettability \cite{Gree78,Raph88,ChWh92} gradient. Especially the Marangoni force
is already used to manipulate droplets, for example in light induced drop
movement \cite{ION00}.
Similar concepts of a directed movement of small amounts of soft matter 
in a given gradient are also realized in models of cell motility \cite{JJP03}. 

However, even more intricate are situations where matter spontaneously 
starts a directed movement in initially homogeneous settings. Small pieces of camphor
that move on a liquid surface by emitting a surfactant
have also been studied for centuries \cite{Vent1799,Toml1869,Rayl1890b,Haya02}.
More recently oil droplets containing volatile additives and interacting droplets
of different volatile oils have been
reported to move on solid surfaces due to the solutal Marangoni effect caused
by evaporation/condensation \cite{CMS64}.
Also intricate is the spontaneous movement of a juxtaposed pair of droplets with different
wetting properties, so called bi-slugs, along a capillary tube
\cite{BiQu00}. This effect was already mentioned by Marangoni
\cite{Mara1871}. 
Another example are drops immersed in a second liquid. If the drops contain 
a soluble surfactant undergoing an isothermal chemical reaction at their 
surface the drops may start to move \cite{RRV94}.
Apart from chemical reactions, drop movement can also be driven by surface
phase transitions \cite{Rieg03,YoPi05_pre}.
The movement is possible because such {\it active} drops change their
surrounding and produce a gradient that drives their motion.
\begin{figure}[tbh]
\includegraphics[width=0.7\hsize]{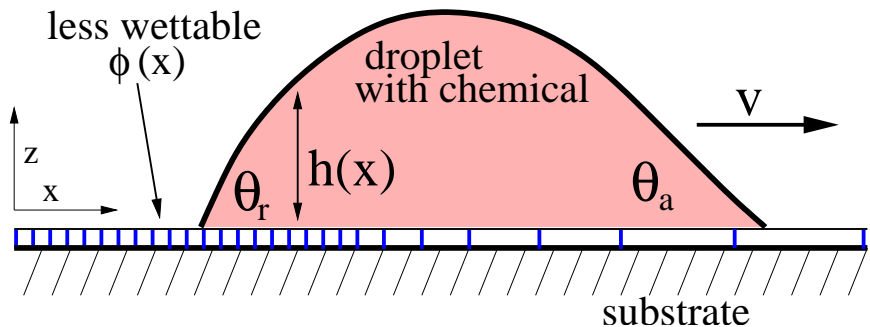}
\caption{Sketch of a right moving droplet driven by a self-produced wettability gradient.
}
\mylab{sketch}
\end{figure}

Recent experiments also found chemically driven running droplets on 
solid substrates \cite{BBM94,DoOn95,LeLa00,LKL02,SMHY05,Sumi05pre,Mage03}. 
In these cases
small droplets of solution are put on partially wettable substrates. 
A droplet changes the substrate by adsorption or desorption of a solute
rendering the substrate underneath the droplet less wettable than the bare substrate.
The radial symmetry still assures an equilibrium position that is, however,
unstable. As a result, fluctuations break the symmetry
and the drop starts to move in a self-sustained manner. In the course of its 
movement it changes the substrate and leaves a less wettable trail behind 
(see Fig.\,\ref{sketch}). We distinguish two types of experiments: In case~I
the substrate is changed in an irreversible way whereas in case~II it
recovers its initial state after the droplet has passed.
%
%
%
The droplet does not rest again and follows a random trajectory until trapping
itself between self-produced less wettable patches (type~I)
\cite{DoOn95,LKL02}. 
Alternatively, a running drop may perform a periodic 
movement (type~II) \cite{SMHY05,Sumi05pre}. 
The droplet often moves with nearly constant velocity for a rather long time
(tens of minutes). However, in all cases the motion ceases when the overall
system reaches thermal equilibrium.
Next, we shortly recall the specific results of different
sets of experiments, yielding running droplets.

In Ref.\,\cite{DoOn95} droplets of $n$-alkanes ($n$-octane and $n$-dodecane)
are used that contain 1,1,2,2-tetra\-hydro\-per\-fluoro\-decyl\-tri\-chlo\-ro\-si\-lane
(CF$_3$(CF$_2$)$_7$(CH$_2$)$_2$SiCl$_3$). The silane molecules
form dense crafted monolayers on silicon or glass and render these surfaces 
hydrophobic. The droplet motion is stalled when no further hydrophilic surface is available. No
limitation due to silane depletion inside the droplet is observed. 
For millimeter sized drops (smaller than the capillary length) droplet velocities
between 1\,mm/s and 10\,cm/s 
are observed, depending on liquid viscosity, silane concentration and droplet 
size. The velocity increases with the silane concentration and the droplet size.
%
%
The reaction rate is estimated to be
1100\,s$^{-1}$mol$^{-1}$. Typical reaction times are given as 0.01--0.2\,s.

Another experiment is performed in a chemically different system \cite{LeLa00,LKL02} using
millimeter-sized droplets of a nonpolar solution of $n$-alkylamine (1\,mM solution of
C$_6$NH$_2$ or C$_{18}$NH$_2$) in  
decahydro-naphthalene ($\gamma=41$\,mJ/m$^2$, $\mu=0.001$\,Ns/m$^2$). Therein,
silicon substrates with microprinted high-energy surfaces are employed, that expose a
dense packing of carboxylic acid functionalities (CO$_2$H). The amines
dissolved in the droplet adsorb
at the substrate and produce a surface of lower energy exposing methyl
groups. The effect of different adsorbates \cite{LeLa00} and reaction kinetics
\cite{LKL02} is investigated. The velocity decreases with increasing droplet size.

The only example of a type~II experiment \cite{SMHY05} features
droplets of oil (5\,mM iodine solution of nitrobenzene saturated with
potassium iodide) on glass substrates immersed 
in aqueous solution of a cationic surfactant (1\,mM stearyl
trimethyl ammonium chloride). In this system the stearyl trimethyl ammonium
ion (STA) absorbs at the glass substrate outside the oil droplet and renders the glass
lyophilic. When the oil droplet moves on top of the coating 
the STA desorbs into the oil. In this way a wettability contrast between front
and back of the moving droplet is created and sustained. In contrast to the
type~I case the substrate recovers its lyophilic state
soon after the droplet has passed, because 'new' STA absorbs at the glass.
The aqueous phase can be seen as an infinite reservoir of adsorbent. 
Movement only finally stops when the oil droplet is saturated with STA.

Similar phenomena can be seen in metallurgic systems where droplets
of liquid metals or alloys react with the metallic substrate, for instance, by
alloying. The layer between the droplet and the substrate may be
less wettable than the bare substrate resulting in the 
migration of reactive islands. This was studied for tin islands on copper
surfaces that move and leave tracks of bronze behind \cite{SBH00}.
Contrariwise, the layer can be more wettable than the bare substrate. This
is typical for the related process of reactive spreading  (also called reactive wetting) 
where a droplet of liquid on a (nearly) non-wettable surface 
starts to spread after forming a more wettable layer underneath \cite{LaEu96,Yost98,WBR98,VMHE99,SCT00,WeGr02}. Variants are possible in
which during spreading the substrate becomes less wettable in the center of
the drop \cite{ZWT98}.
However, reactive spreading processes do normally not result in running
droplets (but see the 'suddenly displaced' droplets in Ref.\,\cite{KRE99}). 

For type~I experiments
an implicit equation for the velocity $v$ of the droplet was derived 
\cite{Brde95,deGe98} from a simple theoretical argument.
Based on a balance of friction force and driving capillary forces one obtains
$v = C \tan \theta^{*} ( 1 - \exp(-r L/v ))$, where 
$r$ is the reaction rate, $L$ the size of the droplet and $C$ a constant.  
The dynamic contact angles at
the advancing and receding ends of the droplet are then assumed to be identical
($\theta^{*}$), i.e.\,the droplet profile is approximated by a spherical cap
with $\theta^{*}$ given by $\cos \theta^{*} = (\cos \theta_e^a + \cos \theta_e^r)/2$.
The static contact angles at the advancing edge $\theta_e^a$ and  
at the receding edge $\theta_e^r > \theta_e^a$ are different due to the chemical
gradient. 
The expression for the velocity is found from a first order
reaction on the substrate that yields chemical concentrations $\alpha_a = 0$ 
and $\alpha_r =  1 - \exp(-r L/v)$ at the respective ends of the 
droplet. 
The expression for the velocity predicts a monotone increase of the
droplet velocity with the droplet length $L$ and the reaction rate
$r$ in line with experimental observations \cite{DoOn95}.  
The droplet velocity in the limiting case of a saturated chemical reaction
was also given in \cite{MiMe97}. 
However, the experiments of Ref.\,\cite{LKL02} show the opposite 
trend; the velocity decreases with increasing drop sizes and effective 
reaction rate. In the framework of Ref.\,\cite{Brde95} the decrease is explained
as a result of the flattening of the drops by gravity.
Related works discuss running
droplets in a random medium \cite{deGe99} and the forced wetting of a plate
immersed into a reactive fluid \cite{deGe97}.

In this paper we propose and analyze dynamical models for self-propelled 
running droplets for both, type~I and type~II experiments. Our models
consist of coupled evolution equations for the
film thickness profile and the substrate coverage. Thereby the 
wettability of the substrate is modeled by a coverage dependent disjoining 
pressure. The different types of experiments are taken into account
by incorporating different reaction terms.
The models are capable of reproducing the different experimentally found regimes. 
A short account of a variant of 
the model for the type~I experiments, i.e.\ for the case of a
irreversibly changed substrate was recently presented \cite{TJB04}. There it
was found that the dynamic contact angles at the advancing and receding ends 
of the droplet are not identical but rather resemble the corresponding static 
ones. Even for a strong driving force the deviation between dynamic and static
angles is only about 10 percent. The bifurcation leading
from sitting to running droplets for a finite reaction rate 
was identified as a drift-pitchfork bifurcation.
Here we analyze the type~I model in more detail and extend our analysis to
the type~II model. 

In section~\ref{model} the type~I and type~II dynamical models are presented.
Results for the two models are discussed in section~\ref{ad} and \ref{addes},
respectively. Stationary running droplets and sitting droplets are
characterized in dependence of the control parameters reaction rate,
droplet volume, diffusion constant of the adsorbate field and desorption rate of the adsorbate (for type~II model only). The resulting
families of solutions and their linear stability
are used to derive phase diagrams that describe the
existence regions for running and sitting droplets.
To further elucidate the drift-pitchfork bifurcation that mediates the transition from sitting to
running droplets also in the type~II model we analyze 
the linear modes that destabilize sitting droplets close to the bifurcation.
Before we conclude in section~\ref{conc}, we use numerical simulations 
to illustrate and analyze periodic droplet motion that is possible in the type~II
model. The simulations are compared with our continuation results 
and the type~II experiment of Ref.\,\cite{SMHY05}. 
%
\section{Model}
\mylab{model}
%
\subsection{Evolution equations}
\mylab{evequ}
%
The evolution of a thin liquid film on a horizontal 
smooth solid substrate is described 
by an equation for the thickness profile $h(x,y,t)$ 
derived from the Navier-Stokes equations using long-wave or lubrication 
approximation \cite{ODB97}
\begin{equation}
\partial_t\,h\,=\,-\nabla\,\cdot\,\left\{\frac{h^3}{3\eta}\, 
\nabla\,p \right\}.
\mylab{film}
\end{equation}
The parameters $\gamma$ and $\eta$ are the surface tension and viscosity
of the liquid, respectively. They define the viscous time scale 
$\tau_v=\gamma L/\eta$, where $L$ is a typical length for the system. 
The change in time of the film thickness profile
equals the gradient of a flow that results as the product of a mobility
and a pressure gradient. The mobility $h^3/3\eta$ corresponds to a parabolic
velocity profile 
\begin{equation}
u(x,z)=\left(\frac{z^2}{2}-z h\right)\, \nabla p
\mylab{velprof}
\end{equation}
within the film. The velocity component orthogonal to the substrate $w$ can be
obtained using the continuity equation
\begin{equation}
\partial_x u + \partial_z w \,=\,0\,.
\mylab{continuity}
\end{equation}
The pressure 
\begin{equation}
p\,=\,\gamma\Delta h\,+\,\Pi(h,\phi)
\mylab{press}
\end{equation}
contains the curvature (or
Laplace) pressure $-\gamma\Delta h$ and the disjoining pressure $\Pi(h,\phi)$. 
The latter comprises effective molecular interactions
between the film surface and the substrate and accounts for the wetting
properties of the substrate \cite{deGe85,Hunt92,Isra92}.
As discussed in detail below, here a mathematically simple
function $\Pi(h,\phi)$ is used that is common in the literature. In
many situations the qualitative outcome only depends on very 
general characteristics
of the disjoining pressure \cite{TNPV02}. The used form 
allows for solutions of Eq.\,(\ref{film}) that represent static (i.e.\ sitting)
droplets with a finite mesoscopic equilibrium contact angle. The disjoining pressure 
is chosen such that the droplets coexist with an ultrathin precursor film.

The evolution of the density of the adsorbed layer on the substrate determines the wettability 
and is modeled by a reaction-diffusion equation for the 
dimensionless field $0<\phi(x,y,t)<1$
\begin{equation}
\partial_t\,\phi\,=\,R(h,\phi) \,+\, d'\,\Delta\phi,
\mylab{rd}
\end{equation}
where the function $R(h,\alpha)$ describes adsorption or desorption 
on the substrate.  The second term
allows for a diffusion of the chemical species along the 
substrate. 
For simplicity we assume that the adsorbate has the same diffusion constant on
the bare and the droplet covered substrate. However, assuming different
constants would not change the presented results qualitatively.
Note that for type~I experiments $\phi$ directly corresponds
to the substrate coverage. However, in a type~II 
experiment the droplet dissolves a more wettable
coating. In this case the coverage corresponds to $1-\phi$. 
Also here one can use Eq.\,(\ref{rd}), however, the
signs of adsorption and desorption term are switched.
With this convention, 
in both cases the wettability decreases with increasing $\phi$.

One can neglect the dynamics of the concentration
field of the chemical in the bulk of the droplet by assuming a fast
equilibration of the solute concentration within the moving droplet as
compared to the reaction at the substrate.
The fast equilibration is caused by diffusion and convective motion within the
droplet whereas the latter is driven 
by the lateral movement of the droplet along the substrate.
It corresponds to the limit of a small Damk\"ohler number 
Da$=r c_0/(D/L)$ giving the ratio of reaction velocity at the substrate 
and diffusion velocity in the droplet \cite{Prob94}. The parameter $r$ is a typical
reaction rate, $c_0$ stands for a typical concentration of the chemical
species in the droplet and $D$ is the diffusion constant in the droplet.
%
\subsection{Reaction term}
\mylab{reaction}
%
Corresponding to the two different sets of experiments that the model shall 
describe we
use two different reaction terms. For type~I the initial coverage of the
substrate is zero (i.e.\ $\phi=0$) 
and we only allow for adsorption underneath the droplet
\begin{equation}
R_1(h,\phi)\,=\,r_{in}\,\xi(h) \,\left(1-\phi\right).
\mylab{reac}
\end{equation}
The reaction saturates at $\phi=1$ because this is the maximal 
possible coverage. The function $\xi(h)$ represents a (smooth) step 
function that approaches one and zero inside and outside the droplet, 
respectively.
For type~II the 'rest' state of the substrate without any droplet is the
fully covered substrate (again corresponding to $\phi=0$, see above section~\ref{evequ}) 
and we allow for desorption underneath the droplet and adsorption 
outside the droplet
\begin{equation}
R_2(h,\phi)\,=\,r_{in}\,\xi(h)\,(1-\phi)\,-\,r_{out}\,[1-\xi(h)]\,\phi.
\mylab{reac2}
\end{equation}
The time scales of the reactions at the substrate inside and outside the
droplet are defined by the effective rate constants
$r_{in}$ and $r_{out}$, respectively. Note that they have the dimension s$^{-1}$. 
The function $\xi(h)$ may be the step function 
\begin{equation}
\xi_1=\Theta(h-h_c), 
\mylab{xi1}
\end{equation}
or the smooth function 
\begin{equation}
\xi_2=\{\tanh[(h-h_c)/\Delta]+1\}/2.
\mylab{xi2}
\end{equation}
The value of $h_c$ is chosen slightly larger than the thickness of 
the precursor film and the maximal drop height is always $h_{max}\gg h_c$.
A small value of $\Delta\ll h_{max}$ 
ensures that the switch between a predominant adsorption reaction
and a predominant desorption reaction occurs over a small film thickness
range.
Note, that $\xi_2\rightarrow\xi_1$ for $\Delta\rightarrow0$.
Changes in the details of the reaction term 
do not affect the results qualitatively. 
%
\subsection{Disjoining pressure}
\mylab{disjpress}
%
For the disjoining pressure we use throughout the present work
\begin{equation}
\Pi(h)=\frac{2 S_l d_0^2}{h^3} + \left(1-\frac{\phi}{g}\right)\,\frac{5 S_s d_0^5}{h^6}
\mylab{pi2}
\end{equation}
where $d_0=0.158$\,nm is the Born repulsion length that defines a lower
cut-off for the film thickness.
The parameter $S_l$ and $S_s$ are the long- and short-range components
of the total spreading coefficient $S=S_l+S_s$ (for  $\phi=0$). We use 
$S_l<0$ and $S_s>0$ corresponding to a destabilizing 
long-range van der Waals and a stabilizing short-range interaction \cite{disjpress}.
For $\phi=0$ the pressure allows 
for drops sitting on a stable precursor film as can be
seen studying the corresponding densities of the excess surface energy $f$, related to the 
disjoining pressures by $\Pi=-\partial_h f$.

The short-range interaction contains the influence of the coating and
crucially influences the static contact angle \cite{Shar93}.  
To account for the varying wettability caused by the different substrate
coverage we let the short-range part of the spreading coefficient $S_s$
depend linearly on the field $\phi(x,y,t)$.
The signs are chosen in a way that for both types of experiments
$g>0$ assures that an increase in $\phi$ corresponds to a
lower wettability, i.e.\ to a larger equilibrium contact angle  $\theta_e$ given by
$\cos\theta_e=S/\gamma+1$ \cite{Shar93}. 
The constant $g$ relates the coverage to the wettability and therefore
defines the magnitude of the possible wettability gradient. 
Note, that our Eq.\,(\ref{pi2}) corresponds to the linear relation
between $\cos \theta_e$  and $\phi$  assumed in Ref.\,\cite{DoOn95,LKL02,Brde95}.

\subsection{Dimensionless equations}
\mylab{dimless}
%
We rewrite Eqs.\,(\ref{film}) to (\ref{pi2})
by introducing scales $3\gamma\eta/l\kappa^2$, $\sqrt{\gamma l/\kappa}$,
and $l$ for $t$, $(x,y)$,  and $h$,
respectively.
Then one obtains 
$2|S_l|d_0^2/l^3$ for the scaled spreading coefficient $\kappa$.
As length scale $l$ we use the value of the film thickness where the local free
energy $f(h)$ has its minimum, i.e., where $\Pi(h)=0$. This gives 
$l=(5S_s/2|S_l|)^{1/3}d_0$ and implies that the ratio 
of the strength of the two antagonistic effective molecular interactions is
only an implicit parameter of the system. The length $l$ also corresponds to
the thickness of the precursor film for a sitting drop on the bare substrate.

Defining the dimensionless overall reaction rate 
$r=3r_{in}\gamma\eta/l\kappa^2$, diffusion constant $d=3d'\eta/\kappa l^2$, 
and ratio of reaction rates $s=r_{in}/r_{out}$,
we obtain from Eqs.\,(\ref{film}) to (\ref{pi2}) 
the dimensionless coupled evolution equations for the thickness profile $h$
and the field $\phi$
\begin{eqnarray}
\partial_t\,h\,&=&\,-\nabla\,\left\{h^3\, 
\nabla\,\left[\Delta h +  \Pi(h,\phi) \right]\right\} \mylab{sys1}\\
\partial_t\,\phi\,&=&\,r\,R(h,\phi)\,+\,d\,\Delta\phi
\mylab{sys2}
\end{eqnarray}
with the disjoining pressure
\begin{equation}
\Pi(h,\phi)\,=\,-\frac{1}{h^3} + \left(1-\frac{\phi}{g}\right)\,\frac{1}{h^6}
\mylab{pi2b}
\end{equation}
and the options for the reaction term
\begin{eqnarray}
R_1(h,\phi)\,&=&\,\xi(h) \,\left(1-\phi\right)
\mylab{r1}\\
R_2(h,\phi)\,&=&\,\xi(h)\,(1-\phi)\,-\,s\,[1-\xi(h)]\,\phi.
\mylab{r2}
\end{eqnarray}
To give an impression of the influence of the coverage we give in
Fig.\,\ref{sitting} droplet profiles for different values of the coverage 
$\phi$. Thereby, $\phi$ is assumed to be independent 
of time and constant along the substrate. Increasing the ratio $\phi/g$
clearly leads to increasing contact angles and decreasing droplet length. 
\begin{figure}[h]
\begin{center}
\includegraphics[width=0.7\hsize]{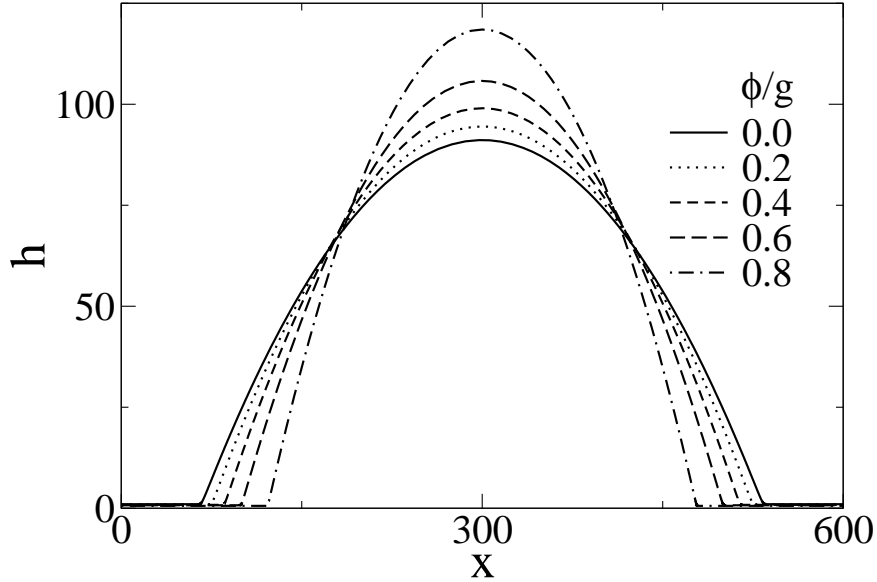}
\end{center}
\caption{Profiles of sitting droplets without reaction for different constant 
values of the wettability measure $\phi/g$ as given in the legend. 
The coverage is assumed to be independent 
of time and constant along the substrate.
The system size is $L=10000$ and all droplets have the volume 
$V\approx{}3\cdot{}10^4$.}
\mylab{sitting}
\end{figure}

In the following we explore the full non-linear behavior of the dynamical
models presented above. The main interest lies in an exploration of
the existence regions of qualitatively different solutions depending on
control parameters like reaction rate $r$, ratio of desorption and adsorption rates
$s$, and droplet volume $V$. It was already shown that a type~I model allows
for three-dimensional
running droplets that move with constant velocity $v$ and droplet shape \cite{TJB04}. Here, we restrict our attention to two-dimensional drops to be
able to explore a large part of the parameter space. 
To determine such droplets we therefore use Eqs.\,(\ref{sys1},\ref{sys2}) 
replacing $\nabla$ by $\partial_x$. Then continuation techniques 
\cite{DKK91,DKK91b,AUTO97} are employed to calculate two-dimensional running 
droplets moving with constant speed.
This is achieved by switching to the comoving frame $x-vt$ and
imposing appropriate boundary conditions. In the comoving
frame running droplets correspond to steady solutions.
Integration in time of the full system is also used in some cases. Details on
all numerical techniques used can be found in the Appendix.
%
%
\section{Results for model~I (only adsorption)}
\mylab{ad}
%
We start with model type~I, where adsorption takes
place only underneath the droplet, i.e.\ we combine the evolution equations for the
film thickness profile (\ref{sys1}) and the field of adsorbate coverage (\ref{sys2}) 
with the reaction term (\ref{r1}) and the disjoining pressure (\ref{pi2b}). 
\subsection{Thickness and coverage profiles}
First we focus on the characterization of the changing solution behavior in 
dependence of the overall reaction rate $r$. Without diffusion along the
substrate ($d=0$) one finds unstable sitting droplets and stable 
running droplets for all $r$. The droplets and the coverage profile move with 
constant velocity $v$ and constant shape. We emphasize that this state corresponds to a subtle
dynamical equilibrium given that coverage profile and droplet 
velocity depend on each other.

Fig.\,\ref{m1prof} shows for two different reaction rates
profiles of moving droplets [(a) and (b)] where the
streamlines correspond to contour lines of the stream function 
\begin{equation}
\psi(x,z)\,=\, \left(\frac{z^3}{2} - \frac{3z^2h}{2}
\right)\,\partial_x p - v z\,,
\mylab{streamfct}
\end{equation}
and indicate the flow in the comoving coordinate system. Note that the velocity fields
are obtained by $(u,w)=(\partial_z\psi,-\partial_x\psi)$.
Also shown are the corresponding profiles for the coverage $\phi$ [(c) and (d)].

The two sets of profiles belong to two qualitatively different regimes 
that are prominently visible in the profiles of the coverage (corresponding to the results
obtained for a different disjoining pressure in \cite{TJB04}).
\begin{figure}[h]
\begin{center}
\includegraphics[width=0.7\hsize]{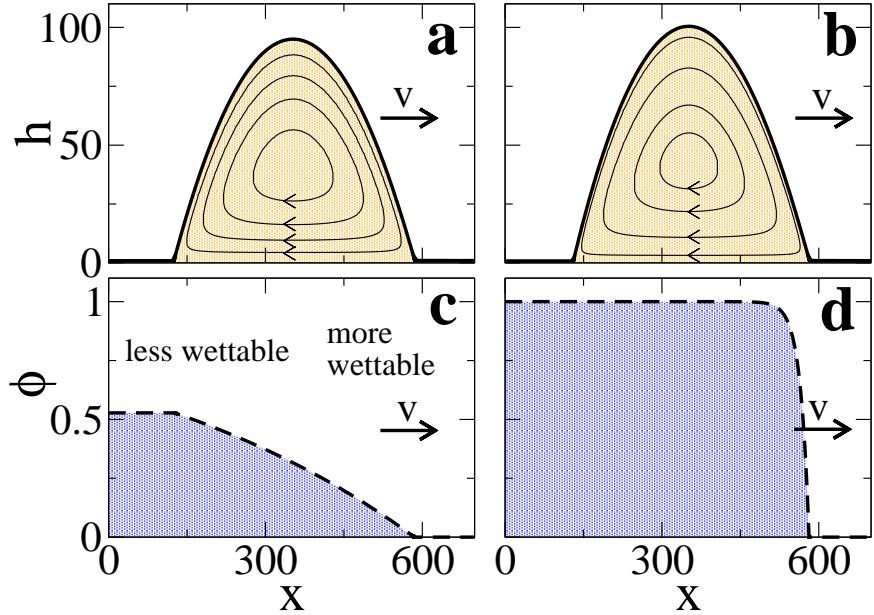}
\end{center}
\caption{Profiles of running droplets in the comoving frame. Shown are (a,b) the
droplet profiles $h$ and (c,d) the substrate coverage $\phi$ for two different reaction rates $r$.
Droplets on the left and right correspond to the non-saturated ($r=10^{-5}$) 
and saturated ($r=0.001$) regime and move with velocities $v\approx 0.006$ and
$v\approx 0.016$, respectively. 
The streamlines plotted in (a) and (b)
indicate the convective motion inside the droplets in the comoving frame and
have the spacing $\Delta\psi=0.025$ and $0.1$, respectively. The
direction of the movement is indicated by arrows.
The remaining parameters are $g=2$, $d=0.001$, $L=10000$, $\bar{h}=3.8$,
$h_c=2.0$, $\Delta{}h=0.2$ and
the droplet volume is $V\approx{}3\cdot{}10^4$.}
\mylab{m1prof}
\end{figure}
\subsection{Dependence on reaction rate}
For low reaction rates [Fig.\ref{m1prof}\,(a) and (c)] 
the coverage starts to increase at the advancing
contact zone and continues to increase up to the receding contact zone where
it is still well below the saturation value of $\phi=1$. We call this the {\it
non-saturated regime}.
For high reaction rates [Fig.\ref{m1prof}\,(b) and (d)] the coverage starts to
increase at the advancing contact zone as in the non-saturated
regime. However, it increases much faster and reaches the saturation value $\phi=1$
already underneath the droplet. At the receding contact zone it is always at
the saturation value. We call this the {\it saturated regime}.
The droplets in the two regimes behave qualitatively different when the reaction rate is
changed as shown in Fig.\,\ref{m1vrcr}.

\begin{figure}[h]
\begin{center}
\includegraphics[width=0.7\hsize]{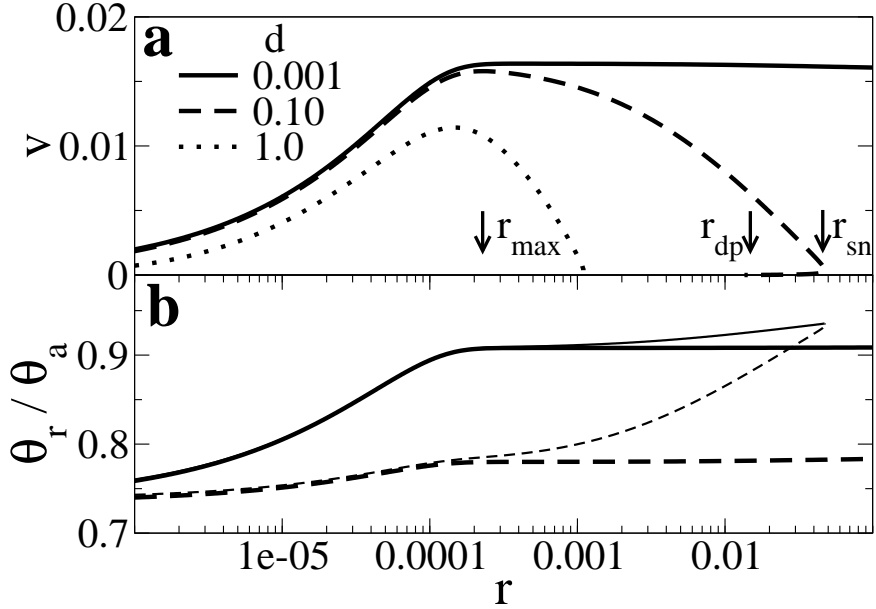}
\end{center}
\caption{Given is a characterization of stationary running droplets in dependence of the
reaction rate $r$ (logarithmic scale) for different strength' of the diffusion of the coverage
$\phi$ along the substrate $d$. Shown are 
(a) the velocity $v$ for different diffusion constants $d$ (see legend); 
and (b) the advancing ($\theta_a$, dashed lines) 
and the receding ($\theta_r$, solid lines) dynamic contact angles for $d=0.001$ (heavy
lines) and $d=0.1$ (thin lines). The arrows with the labels $r_{max}$,
$r_{dp}$ and $r_{sn}$ in (a) indicate the locations for the maximum velocity,
drift-pitchfork bifurcation and the saddle-node bifurcation, respectively, for
$d=0.1$. The remaining parameters are as in Fig.\,\ref{m1prof}.}
\mylab{m1vrcr}
\end{figure}
For low reaction rates in the non-saturated regime 
the coverage does not reach the saturation
value $\phi=1$ at the rear of the droplet. This implies that the driving wettability gradient
between front and rear of the droplet has not yet reached its maximally possible value.
Therefore an increase in the reaction rate leads to a steeper increase
in the spatial profile of the coverage underneath the droplet. 
In consequence a larger wettability 
gradient results implying a larger velocity. However, again it is subtle to determine
the dynamic equilibrium. First, the higher velocity reduces the contact
time of the droplet with a given point of the substrate. Second, it also reduces
the length of the droplet (see below). Both effects restrict the
growth of the wettability gradient.
The resulting increase of the velocity with increasing reaction rate can be
seen in Fig.\,\ref{m1vrcr}\,(a) for $r<0.0001$. Note that it does only weakly
depend on the diffusion along the substrate.

Fig.\,\ref{m1vrcr}\,(b) shows the corresponding dynamical 
contact angles at the advancing and receding edges of the moving droplets. 
Below $r\approx0.0001$ both angles increase, implying a decrease in the length
of the droplet (that has constant volume). Because in the non-saturated regime
the coverage in the contact zone at the advancing edge does  
effectively not depend on $r$, the increase in the advancing contact angle is caused by
the larger velocity only. Seeing the full dynamic contact angle at a moving contact line as a
superposition of a static (or equilibrium) part and a dynamic part, 
one can say that only the dynamic part of the advancing contact angle increases 
with $r$. The increase at the front is in line with the increase with velocity 
found for sliding drops on an incline (see for example Ref.\,\cite{Thie02}).
However,  at the receding edge the coverage in the contact
zone depends strongly on $r$. 
Therefore the rather strong increase in the receding contact angle with
increasing $r$ is caused by
changes in both, the static and the dynamic part. For a larger $r$ the
wettability at the back is smaller, i.e.\ the 'local equilibrium 
contact angle' increases. However, the dynamic receding contact angle
normally decreases with increasing velocity \cite{Thie02}. This implies that
the increase in the receding angle found here results from an increase
of the static part only. It is even counteracted by a decrease of the dynamic part. Also
in the contact angles, a relatively small $d$ has no visible influence in the 
non-saturated regime.

The maximum wettability gradient, i.e.\ driving force, 
and therefore the maximum droplet velocity is
reached for the bare substrate ($\phi=0$) at the advancing edge 
and maximum coverage ($\phi=1$) at the receding edge. The point of largest
velocity [Fig.\,\ref{m1vrcr}\,(a)] corresponds to the point of the largest
difference between the receding and advancing contact angles
[Fig.\,\ref{m1vrcr}\,(b)]. We call the corresponding reaction rate $r_{max}$.

Next we discuss the behavior in the saturated regime ($r>0.0001$ in
Fig.\,\ref{m1vrcr}). 
For larger reaction rates the coverage at the rear of the droplet remains
at its saturation value. However, the driving force and in consequence the
velocity decrease slightly with increasing $r$ (cf.\ also Ref.\,\cite{TJB04}). 
Note, that for small $d$ in Fig.\,\ref{m1vrcr}\,(a) this is barely
visible. The decrease from $r_{max}$ to $r=0.1$ corresponds to about 2\% of
the maximal velocity.

As detailed next this slight decay is caused by the dynamics in the advancing
contact zone. In the saturated regime the time scale of the reaction is 
short compared to the one of the droplet movement. Therefore the coverage
increases very steeply underneath the droplet 
leading to an elevated coverage already in the
advancing contact zone (see Fig.\,\ref{m1prof}\,b). The resulting decrease of 
the overall wettability gradient felt by the droplet causes the slight decrease of the velocity.
In this regime the slight increase of the advancing contact angle results from an
increase of the static part (because of the decreasing wettability) that overcomes
a decrease of the dynamic part (decreasing velocity). 
The slight increase of the receding angle comes from the dynamic part only.

We emphasize here that Fig.\,\ref{m1vrcr}\,(b) clearly shows that
the advancing contact angles are always smaller than the receding
ones. As was already shown in Ref.\,\cite{TJB04} also here the differences between
the static and the dynamic contact angles at the front and the rear, respectively, 
are an order of magnitude smaller than the difference between
the two static (or the two dynamic) contact angles. 
\subsection{Phase diagrams}
\subsubsection{Influence of Diffusion}
\begin{figure}[h]
\begin{center}
\includegraphics[width=0.7\hsize]{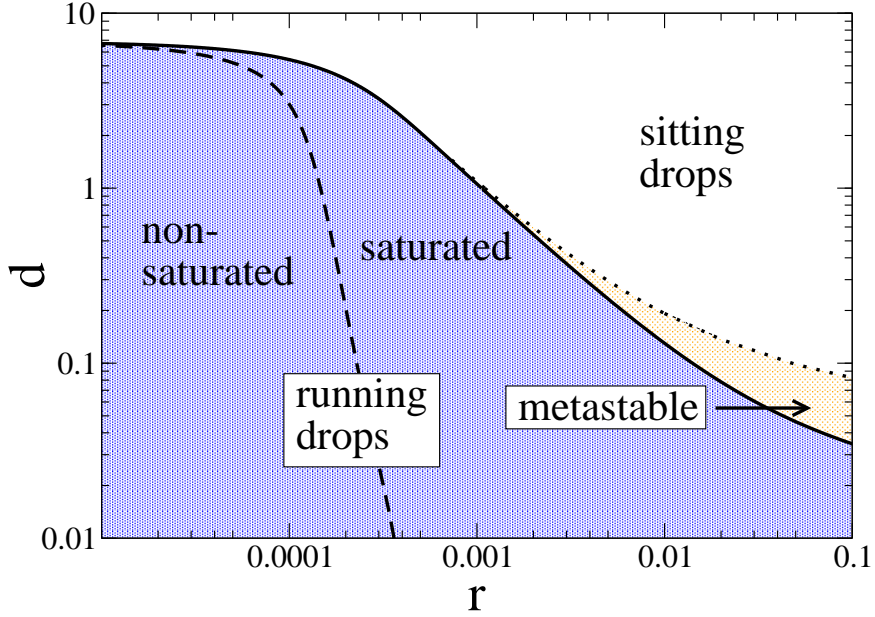}
\end{center}
\caption{Existence of running and stationary droplets depending on the
reaction rate $r$ and the diffusion constant $d$. The solid line indicates a
drift-pitchfork bifurcation where a running droplet solution branches off a
stationary droplet solution. The dotted line indicates a saddle-node bifurcation where two
running droplet solutions merge. Shown is also the transition
between non-saturated and saturated regime (dashed line). The remaining 
parameters are as in Fig.\,\ref{m1prof}.}
\mylab{m1phase}
\end{figure}
Let us finally discuss the influence of diffusion along the substrate on the 
dependencies shown in Fig.\,\ref{m1vrcr}. The result without diffusion (not shown)
coincides nearly perfectly with the shown curves for $d=0.001$. Only the decay
in the saturated regime is slightly slower. We mention here that the decay
also depends slightly on the used cutoff $h_c$ 
(cf.\ Eq.\,(\ref{xi1}) and Ref.\,\cite{TJB04}). Increasing the diffusion has a rather
small influence on the non-saturated regime but changes the saturated regime
even qualitatively. The velocity decreases with the reaction rate because 
the adsorbate produced underneath the droplet close to the advancing edge diffuses
onto the bare substrate ahead of the moving droplet. This effectively reduces
the wettability there, implying a slower movement. In fact, running droplets
cease to exist above a reaction rate $r_{dp}$ where the velocity drops to zero
again, corresponding to a supercritical bifurcation for large diffusion 
(see curve for $d=1.0$).
The bifurcation structure can, however, be more involved. For a smaller
$d=0.1$ one finds that the branch of running droplets joins the branch of
sitting drops in a subcritical bifurcation at $r_{dp}$, 
i.e.\ decreasing $v$ the running
droplet branch first turns back in a saddle-node bifurcation at $r_{sn}$ and
then joins the sitting droplet branch at $r_{dp}$. We indicate the definition
of $r_{sn}$ and $r_{dp}$ in Fig.\,\ref{m1vrcr}\,(a) using the curve for
$d=0.1$ as example. However, the subcritical part of the branch is difficult
to see. We refer the reader to the discussion of model~II in 
section~\ref{addes} for more details. Between $r_{dp}$ and $r_{sn}$ both the large
velocity running droplet and the sitting droplet are linearly stable and
correspond to metastable states. The branch emerging from the subcritical
bifurcation is linearly unstable
and corresponds to threshold solutions that separate the two
stable solutions.
The prominent features ($r_{max}$, $r_{dp}$ and $r_{sn}$) 
of the solution branches shown in Fig.\,\ref{m1vrcr}
can be used to determine phase diagrams by continuation
techniques \cite{AUTO97}. The phase diagrams show existence
regions for the different types of solutions in the parameter space spanned,
for instance, by the reaction rate and the diffusion constant (see
Fig.\,\ref{m1phase}). 
To this end we followed the maximum of the 
$v(r)$ dependence at $r_{max}$, which marks the transition between the non-saturated and saturated
regime for running droplets (dashed line in Fig.\,\ref{m1phase}). 
Continuation of the loci of the saddle-node bifurcation at $r_{sn}$ and the
drift-pitchfork bifurcation at $r_{dp}$ gives the border of
the existence region for stable running droplets (dotted line) and stable sitting droplets
(solid line), respectively. The region between the two latter borders
corresponds to a coexistence region where running and sitting droplets are metastable.
Note, that unstable sitting drops do also exist everywhere in the existence region
of the running droplets. 
Such sitting droplets are steady solutions of the governing
equations. However, because they are unstable, even infinitely small
perturbations will grow with a characteristic rate. In consequence the droplets start to move 
and adopt the shape and speed of the stable running droplet solution. 
\subsubsection{Influence of volume}
Practically, it is difficult to change the reaction rate over orders 
of magnitude. Experiments normally only cover a much smaller range
\cite{DoOn95,LKL02}. 
Nevertheless, the analysis of the $v(r)$ dependence
(Fig.\,\ref{m1vrcr}) for droplets of fixed volume gives a very good
characterization of the general system behavior. An experimentally important
parameter is the size or volume of the droplet. The velocity may increase
\cite{DoOn95} or decrease \cite{LKL02} with increasing volume depending on 
the parameter regime \cite{TJB04}. We show in Fig.\,\ref{vLI} the dependency
of velocity on droplet volume (logarithmic scale) for a variety of reaction rates.
Also there one can well distinguish a saturated and a non-saturated regime.
In the non-saturated regime the droplet velocity increases with
increasing size, whereas in the saturated regime it decreases. 
For small droplets one always finds a non-saturated regime whereas droplets of
a very large size are always in a saturated regime (in Fig.\,\ref{vLI} the
latter is not yet reached for the curve with $r=10^{-5}$).
The explanation for this behavior is closely related to the one given for the $v(r)$
dependence. For small droplets the reaction does not reach saturation until
the rear of the droplet has passed. This implies that an increase in droplet
size gives more time for the reaction leading to a larger value of $\phi$ at
the back and, in consequence, to a larger velocity. 

For very large droplets the reaction has enough time to reach saturation
($\phi=1$) at the rear, i.e.\ an increase in size does not increase the
driving wettability gradient. It may even decrease slightly due to an increase
of $\phi$ in the contact zone at the advancing edge (see above).
However, a larger droplet has a larger viscous dissipation leading to a
decreasing velocity with increasing size. Inspecting  Fig.\,\ref{vLI} shows
that for $r\ge{}5\cdot{}10^{-4}$ and $V=30\,000$ (i.e.\ the volume used in 
Fig.\,\ref{m1vrcr}) one is well in the saturated regime.

\begin{figure}[h]
\begin{center}
\includegraphics[width=0.7\hsize]{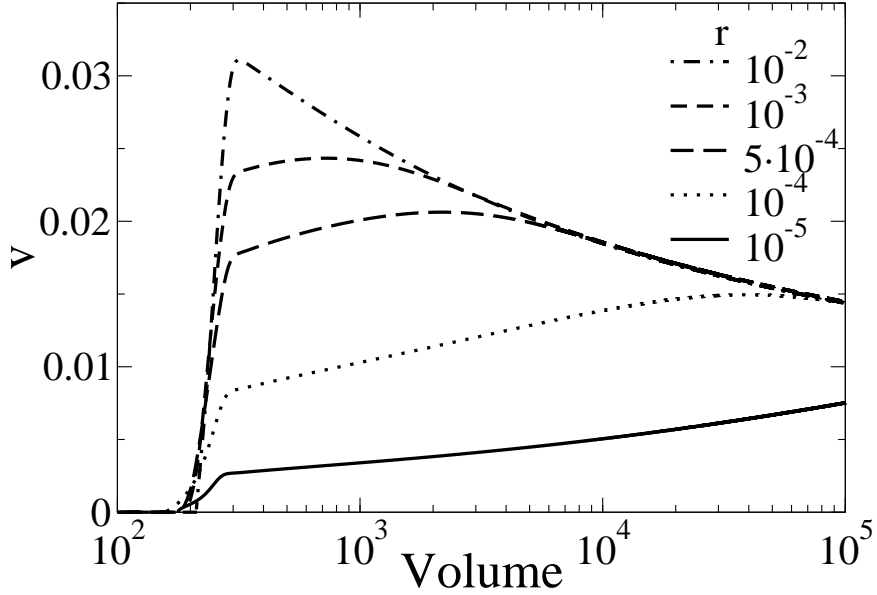}
\end{center}
\caption{The transition between non-saturated and saturated regime is most
clearly seen in the dependence of the velocity $v$ on drop size, i.e.\ droplet
volume $V$ (logarithmic scale). Curves are plotted for different reaction rates $r$ (see
legend). Remaining parameters are as in Fig.\,\ref{m1prof}.}
\mylab{vLI}
\end{figure}

\begin{figure}[h]
\begin{center}
\includegraphics[width=0.7\hsize]{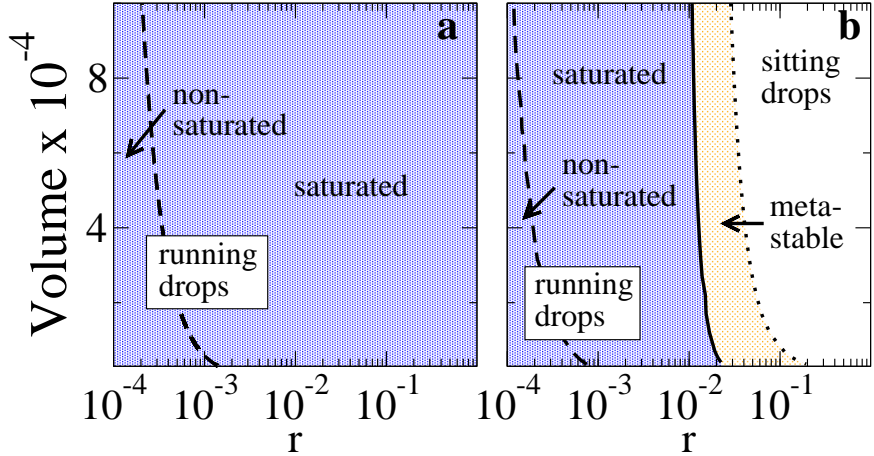}
\end{center}
\caption{Existence of running and stationary droplets depending on the
reaction rate $r$ and the droplet volume $V$ (in units of $10^4$) for two diffusion constants in
the $\phi$-field $d=0.001$ (a) and $d=0.1$ (b). The solid and dotted lines
indicate a drift-pitchfork and a saddle-node bifurcation, respectively. The dashed line indicates 
the transition between non-saturated and saturated regime.
Remaining parameters are as in Fig.\,\ref{m1prof}.}
\mylab{m1existvol}
\end{figure}

The volume $V_{max}$ where the maximal velocity is obtained can be followed in
the space spanned by droplet volume and reaction rate. The obtained existence 
region for non-saturated and saturated running droplets are shown in
Fig.\,\ref{m1existvol}. 

Having studied the type~I model describing experiments where the passing
droplet irreversibly changes the substrate we next report on results for
our type~II model where the substrate recovers. In contrast to type~I this
allows for the description of a periodic droplet movement as experimentally 
observed in Ref.\,\cite{SMHY05}.
%
\section{Results for model~II (adsorption and desorption)}
\mylab{addes}
%
This section is devoted to the type~II model that extends and generalizes 
the model of type~I discussed above. Model~II
accounts for experimental situations where both, the droplet and its
surrounding medium are able to change the substrate by adsorption or
desorption. In this way a moving droplet makes the substrate less wettable, 
but after the droplet has passed the substrate may relax to its
initial state.
Such systems are modeled by extending the reaction kinetics 
for the $\phi$-field (\ref{r1}) by an additional
desorption term, i.e.\ we combine the evolution equations for the
film thickness profile (\ref{sys1}) and the adsorbate field (\ref{sys2}) 
with the reaction term (\ref{r2}) and the disjoining pressure (\ref{pi2b}). We
will refer to this set of equations as type~II model.
The reaction term (\ref{r2}) is chosen such that adsorption and desorption of $\phi$ take
predominantly place underneath  and outside the droplet, respectively.

\subsection{Thickness and coverage profiles}
\begin{figure}[hbt]
\begin{center}
\includegraphics[width=0.7\hsize]{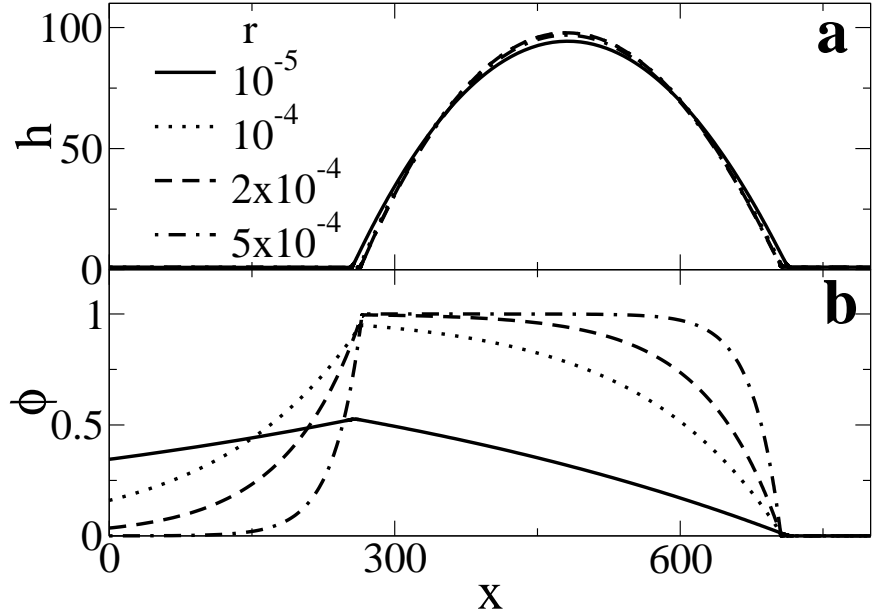}
\end{center}
\caption{Results of model~II incorporating adsorption and desorption. 
Shown is the influence of the reaction rate $r$ (see legend) for 
(a) the profiles of moving droplets and (b) 
the corresponding profiles in the $\phi$-field. The drop profiles at
different $r$ differ only slightly by visual inspection. The differences are
more prominent in the $\phi$-profiles. Parameters are $s=1$, $h_c=2$,
$\Delta{}h=0.2$, $g=2$, $d=0.001$, $V\approx{}30000$ ($L=10000$, $\bar{h}=3.8$).}
\mylab{m2profr}
\end{figure}
As above for model~I, we use continuation techniques to 
calculate running droplet solutions of 
model~II as solutions stationary in a comoving frame.
Examples of the resulting droplet profiles along with the profiles of 
the $\phi$-field are shown for a very small diffusion $d$ 
in Figs.\,\ref{m2profr} and \ref{m2profs}. 
Thereby, Fig.\,\ref{m2profr} displays the results for several values of the
reaction rate $r$ for a fixed ratio $s$ of the desorption to adsorption rates.
The drop profiles change only slightly but the $\phi$-profile undergoes
prominent changes. As in model~I there is practically no $\phi$-field
in front of the droplet ($\phi=0$). The field increases (i.e.\ the wettability
decreases) underneath the droplet as in model~I, but in contrast the 
$\phi$-field decreases behind the droplet due to the desorption. 
The concentration of the $\phi$-field at the rear of
the droplet is increasing with $r$ until it reaches saturation. Because $r$ is
the overall reaction rate also the desorption of $\phi$ behind the 
droplet becomes faster, i.e.\ the tail of the $\phi$-field becomes shorter.  

\begin{figure}[hbt]
\begin{center}
\includegraphics[width=0.7\hsize]{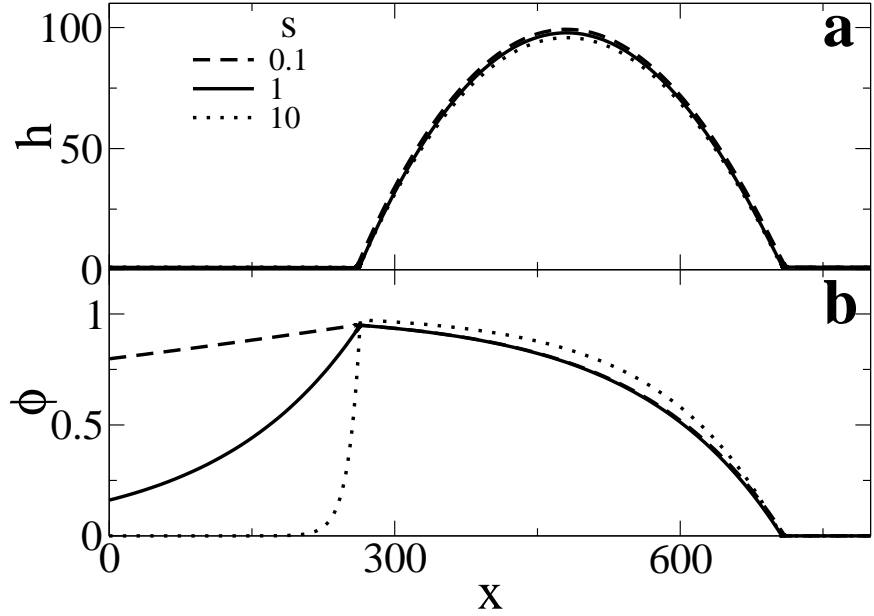}
\end{center}
\caption{Shown is the influence of the ratio $s$ of desorption and
reaction (see legend) on (a) the profiles of moving droplets and (b) 
the corresponding profiles in the $\phi$-field for model~II. The drop profiles at
different $s$ differ only slightly by visual inspection. The differences are
prominent in the tail of the $\phi$-profile behind the droplet. 
$r=10^{-4}$ and the remaining parameters are as in Fig.\,\ref{m2profr}.}
\mylab{m2profs}
\end{figure}

Fig.\,\ref{m2profs} displays droplet and $\phi$ profiles for a fixed reaction
rate $r$ but different values for the desorption to adsorption ratio $s$. The
droplet shape and the form of the $\phi$-field underneath the droplet are 
almost independent of $s$ as one would expect. However the decay 
of the elevated $\phi$ value occurring behind the droplet is strongly 
influenced. For a small $s$, i.e.\ slow desorption as compared to adsorption,
$\phi$ decays slowly approaching qualitatively the behavior of model~I.
Increasing $s$  reduces drastically the length of the $\phi$ tail behind the
droplet. The length of the tail is very important when studying the periodic
movement of a droplet on a finite stripe-like substrate (see below).    
\subsection{Phase diagrams}
In the following the existence regions of running and sitting droplets are
determined in their dependence on the control parameters 
reaction rate, desorption to
adsorption ratio, diffusion constant and droplet volume.
This is done along the lines explained in section~\ref{ad}. Continuation gives
branches of stationary solutions depending on one control parameter. 
On these branches special points that separate 
qualitatively different behavior are identified and followed in the space 
of parameters. The linear stability of the stationary solutions is also
determined. Details on the used techniques can be found in the Appendix.

\begin{figure}[hbt]
\begin{center}
\includegraphics[width=0.7\hsize]{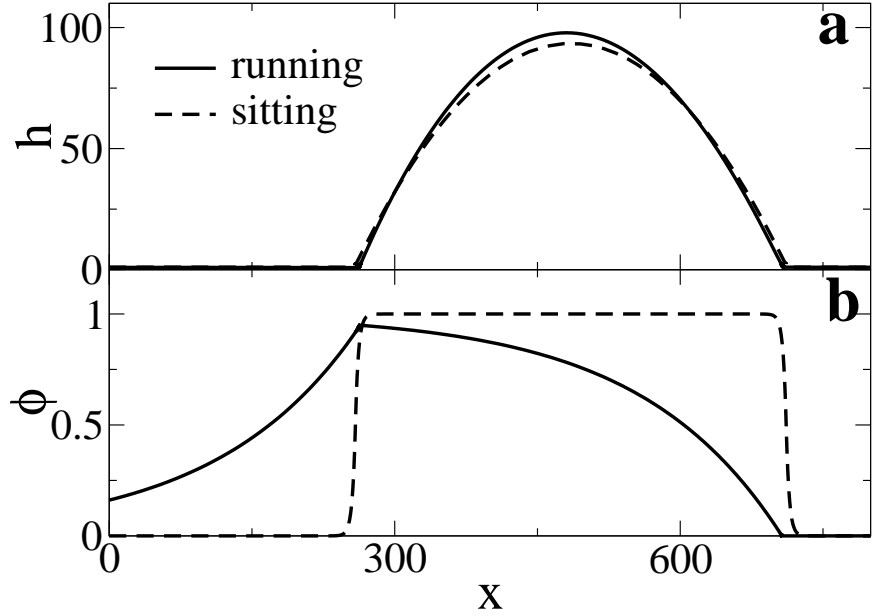}
\end{center}
\caption{
Shown are profiles of (a) the film thickness and (b) the $\phi$-field for
running and sitting droplets (see legend) for model~II. 
The drop profiles differ only slightly by visual inspection. The differences are
more prominent in the $\phi$-profiles. $r=10^{-4}$ and the remaining parameters
are as in Fig.\,\ref{m2profr}.}
\mylab{sitdrop}
\end{figure}
\subsubsection{Influence of the desorption/adsorption ratio}
\begin{figure}[h]
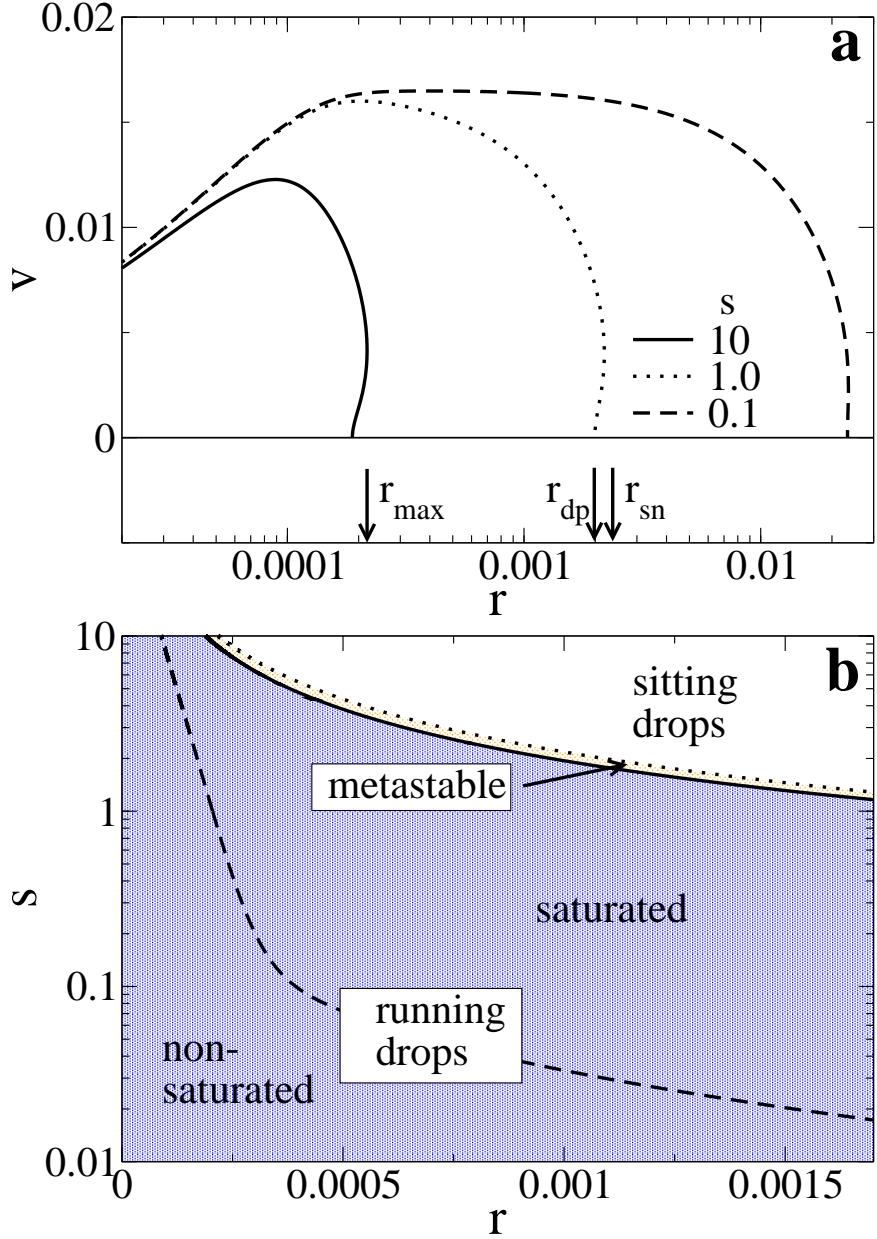

\begin{center}
\includegraphics[width=0.7\hsize]{m2phases_a.eps}\\
\includegraphics[width=0.7\hsize]{m2phases_b.eps}
\end{center}
\caption{(a) Velocity of running droplets depending on the reaction rate $r$. The
branch of running droplets emerges from zero reaction rate and undergoes a
subcritical drift pitchfork bifurcation at finite $r>0$. The reaction
rate-velocity curve is shown for three different values of the ratio between
desorption and adsorption $s$. The labels $r_{max}$,
$r_{dp}$ and $r_{sn}$ indicate the locations for the maximum velocity,
drift-pitchfork bifurcation and the saddle-node bifurcation, respectively, for
$s=1.0$. (b) Existence of running and sitting droplets depending on the
reaction rate $r$ and the ratio between desorption and adsorption $s$. Shown is also the boundary
between non-saturated and saturated regime (dashed line). The remaining parameters are
as in Fig.\,\ref{m2profr}.
\mylab{m2vrs}
}
\end{figure}
Studying first the influence of $r$ and $s$ 
we present in Fig.\,\ref{m2vrs}\,(a) and
(b) branches of stationary solutions in dependence of $r$ for different values
of $s$ characterized by their velocity and the resulting existence regions
of running and sitting droplets in the $r-s$ parameter plane, respectively.
Beside the moving droplets there exist sitting droplets (steady states)
for all values of $r$ and $s$. They are symmetrical with respect to the 
droplet maximum and may be stable or unstable. 
An example is shown in Fig.\,\ref{sitdrop}.

The curve for $s=1$ in Fig.\,\ref{m2vrs}\,(a) is used to better illustrate 
special values of the reaction rate introduced above in section~\ref{ad}.
They mark the loci of the maximum of the $v(r)$ dependence at $r_{max}$,
of the saddle-node bifurcation at $r_{sn}$ and of the
drift-pitchfork bifurcation at $r_{dp}$. As before, continuation in $r$ 
of these loci when changing $s$ gives the phase diagram  
Fig.\,\ref{m2vrs}\,(b). In particular one obtains
the border between the non-saturated and saturated regime for running droplets 
($r_{max}$), the border of the existence region for stable running droplets
($r_{sn}$), and  the border of the existence region for 
stable sitting droplets ($r_{dp}$). In the small region where both, running  
and sitting droplets are stable, one finds metastability. There noise, for
instance, in the form of substrate inhomogeneities may lead to intermittent
droplet movement. The sitting droplets existing in the gray shaded area of 
Fig.\,\ref{m2vrs}\,(b) are all unstable. 

Inspection of Fig.\,\ref{m2vrs} reveals the rather strong influence of the 
desorption/adsorption ratio $s$ especially on the transition from the
non-saturated to the saturated regime and the slowing down in the saturated
regime. The extension in $r$ of the saturated regime is drastically reduced 
with increasing $s$. For decreasing $s$ the maximum of the $v(r)$ curve slowly
transforms into a plateau and reaches for $s=0$ the form typical for model~I. 
Comparing the strong decrease of the velocity with increasing $r$ in the 
saturated regime [Fig.\,\ref{m2vrs}\,(a)] to the slight decrease found 
for $s=0$ (model~I, Fig.\,\ref{m1vrcr}\,(a), curve for $d=0.001$) poses the
question why $s$ has such a strong influence. 

The influence of $s$ parallels the influence of the diffusion $d$ in model~I.
There for large $d$ the adsorbate diffuses in front of the advancing droplet
rendering the  substrate there less wettable. Thereby it decreases the overall
wettability gradient between front and rear of the droplet. Here, for large
$s$ the $\phi$-field is removed behind the rear of the droplet rendering the
substrate there more wettable. That also implies a reduction of the overall
wettability gradient.
\subsubsection{The drift-pitchfork bifurcation}
To elucidate the mechanism of the transition between moving and sitting drops
we focus for a moment on the drift-pitchfork bifurcation at $r_{dp}$ that separates
the metastable and the running droplet region. This type of bifurcation
is well known from reaction-diffusion 
(see, for instance, \cite{KTB92,KrMi94,HaMe94,OBSP98,Boed03} and references
therein) and hydrodynamic systems
(see, for instance, \cite{PrJo88,CGG89,KnMo90,RiPa92} and references
therein), where it mediates the transition between steady and travelling structures.
In our system it breaks the reflectional symmetry of the sitting droplets leading
to moving asymmetric droplets and acompanying travelling asymmetric 
adsorbate profiles.
At the bifurcation a real eigenvalue switches sign, i.e.\
at $r_{dp}$ the velocity of the moving drops is zero. Beyond the bifurcation
sitting droplets become unstable and start to move slowly.  
That makes it unlike a Hopf bifurcation (associated with the zero crossing of the real
part of a pair of complex eigenvalues) where the travelling structure has a
finite velocity at the bifurcation (for waves on flowing thin liquid films see
the discussion in Ref.\,\cite{ThKn04}). The bifurcation may be
subcritical or supercritical (see Figs.\,\ref{m1vrcr}, \ref{m2vrs} and
\ref{m2vrd}) and in consequence the bifurcating branch corresponds to unstable
or stable moving drops. Approaching the bifurcation their respective 
velocity goes to zero as
$\sqrt{|r-r_{dp}|}$ providing an unambiguous signature of the 
drift-pitchfork bifurcation.

The basic mechanism of the drift-pitchfork bifurcation is connected to the
behaviour of the neutral (or Goldstone) mode related to the translational
symmetry of the system.
This mode with eigenvalue zero is
obtained by analyzing the linear stability of the stationary solutions. 
In general, each continuous symmetry is related to such a neutral mode.
In a sense, the neutral modes are the modes that are 'closest' to zero. This implies that 
a perturbation or modulation of these modes in an additional 
degree of freedom (if existing) gives modes that are probable 
candidates to cross zero and become instability modes. 
For instance, the transversal (fingering)
instability of a liquid front results from a transversal
modulation of the longitudinal translational neutral mode \cite{ThKn03}.
Also one of the two coarsening modes of two liquid droplets is the combination
of translational neutral modes of the individual droplets directed in opposite
directions \cite{MPBT05}.

Here, the drift instability is associated with a mode representing a 
relative shift between the translational modes of the height and the $\phi$ 
profiles. Right at the bifurcation this mode corresponds exactly to the 
translational neutral mode. Although elsewhere 
one can still identify the translational 
mode when only looking at the sub-mode for the height profile {\it or} the one
for the $\phi$ profile, looking at the complete mode one realizes that the
relative weight of the two sub-modes is shifted in favor of one of them. This
corresponds to the introduction of a relative shift between the 
two fields. The relative shift breaks the overall reflection symmetry 
and leads to the movement of the drops.

\begin{figure}[h]
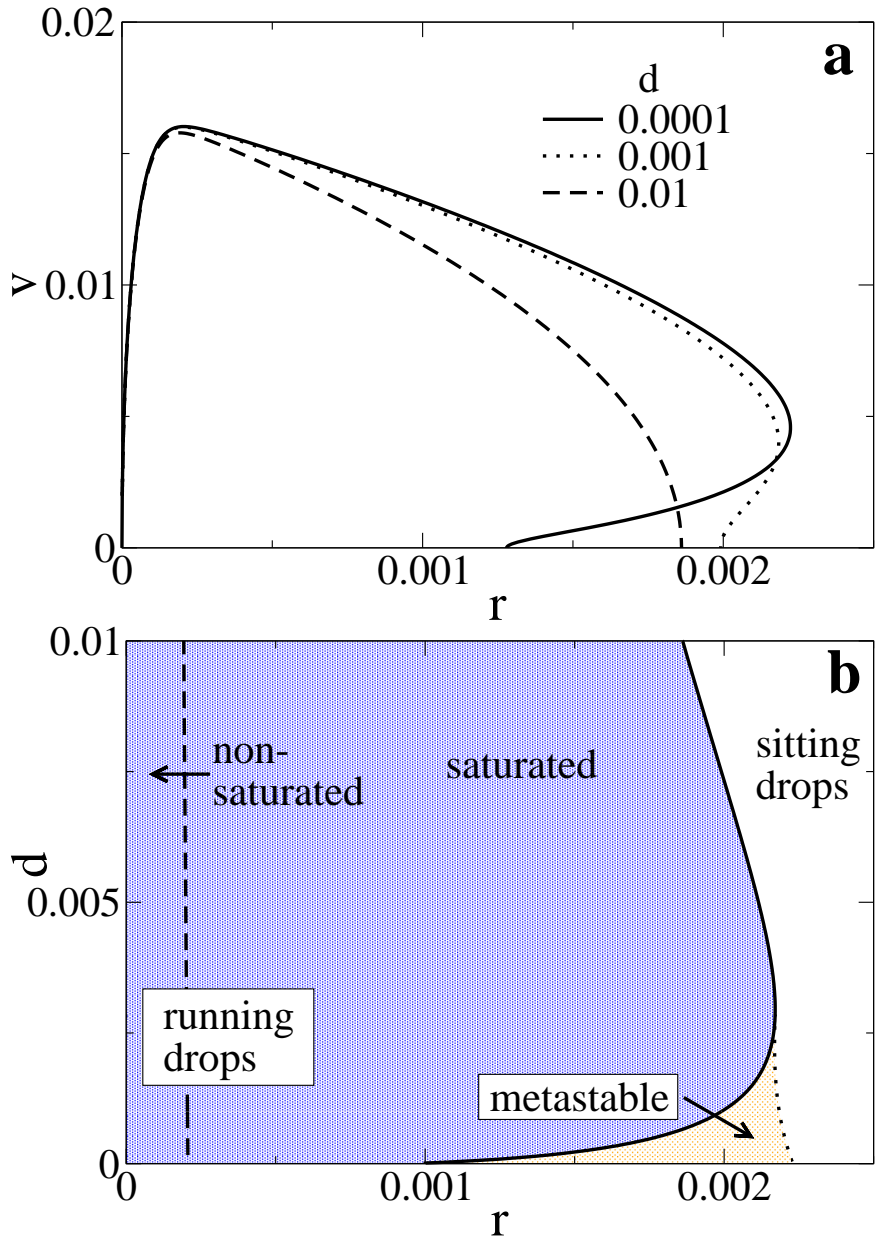

\begin{center}
\includegraphics[width=0.7\hsize]{m2phased_a.eps}\\
\includegraphics[width=0.7\hsize]{m2phased_b.eps}
\end{center}
\caption{(a) Velocity of running droplets depending on the reaction rate $r$. The
branch of running droplets emerges from zero reaction rate and undergoes a
subcritical drift pitchfork bifurcation at finite $r>0$.The reaction
rate-velocity curve is shown for three different values of the diffusion
constant in the $\phi$-field $d$. (b) Existence of running and stationary droplets depending on the
reaction rate $r$ and the diffusion constant of the $\phi$-field
$d$. Shown is also the transition
between non-saturated and saturated regime (dashed line).
Remaining parameters are as in Fig.\,\ref{m2profr}.
\mylab{m2vrd}
}
\end{figure}
\subsubsection{Influence of diffusion}
Next we discuss the influence of $r$ and $d$ presenting in Fig.\,\ref{m2vrd} 
dependencies of droplet velocities on reaction rate for different diffusion
constants and the resulting phase diagram in the $r-d$ plane for a fixed
desorption/adsorption ratio $s$.
One notes first that the strength of diffusion has nearly no 
influence on the non-saturated regime and the transition value $r_{max}$
[the dashed line in Fig.\,\ref{m2vrd}\,(b) is practically vertical].
Note that Fig.\,\ref{m2vrd} focuses on relatively small diffusion. 
For large diffusion the droplets will also stop as shown for model~I
in Fig.\,\ref{m1phase}.
However, the saturated regime does depend on $d$ quantitatively
as well as qualitatively. For fixed $r$, with increasing $d$ the stable running droplets 
become slower. In parallel its existence range in $r$ shrinks slowly.
As in model~I this behavior is mainly caused by the diffusion of the $\phi$-field in
front of the advancing droplet. There it increases the coverage 
thereby reducing the overall wettability gradient, i.e.\ slowing down the droplets.
The qualitative change concerns the character of the drift-pitchfork
bifurcation. With increasing $d$ it becomes less subcritical and at a critical
$d_c$ it becomes supercritical, i.e.\ there exists no coexistence region for 
sitting and running drops any more. For small diffusion $d<d_c$ 
the existence range of stable sitting drops shrinks with increasing $d$
whereas for larger diffusion $d>d_c$ it grows.

\begin{figure}[h]
\begin{center}
\includegraphics[width=0.7\hsize]{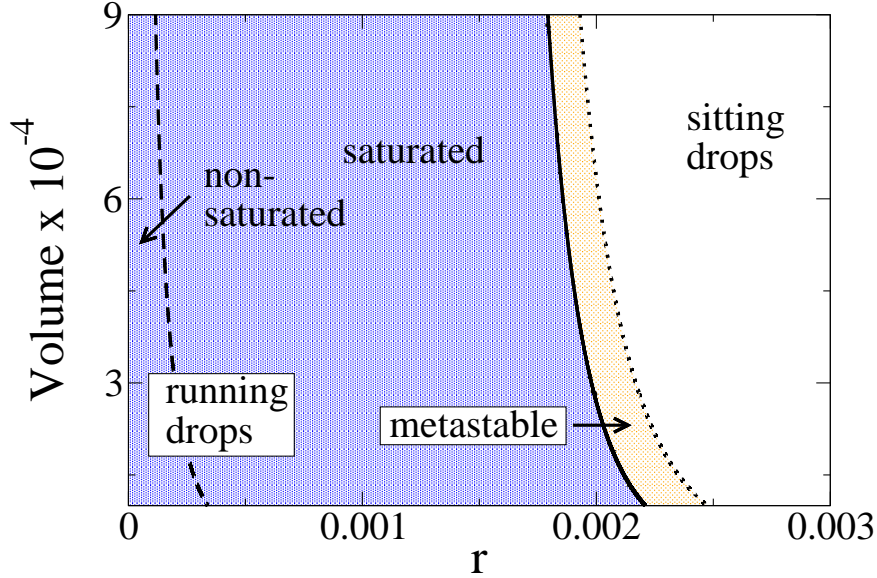}
\end{center}
\caption{Existence of running and stationary droplets depending on the
reaction rate $r$ and the droplet volume $V$ (in units of $10^4$). 
Shown is also the transition
between non-saturated and saturated regime (dashed line). Remaining parameters are as in Fig.\,\ref{m2profr}.}
\mylab{m2vrh}
\end{figure}
\subsubsection{Influence of volume}
Finally, in Fig.\,\ref{m2vrh} we present the phase diagram for the dependence 
on reaction rate and droplet volume for fixed desorption/adsorption ratio $s$
and diffusion constant $d$.
The locations of the saddle-node, the drift-pitchfork bifurcation and the velocity
maximum are all shifted slightly towards smaller $r$ when increasing the droplet
volume. The range in $r$ of the non-saturated regime shrinks slightly, but the
range of the saturated regime and the metastable region remain practically 
constant, they are only shifted towards smaller $r$.

Obviously the two contact regions and the substrate outside the droplet 
are not affected by a change in
the droplet volume. However, an increase in droplet size increases the viscous forces and therefore
stalls the droplet movement at smaller reaction rates. 
\subsection{One-dimensional numerical simulations}
Finally, we employ numerical simulations of running droplets and show that the
type~II model is able to describe different experimentally found modi of periodic 
droplet movement \cite{SMHY05,Sumi05pre}.
To perform the simulations we use the routine 'd02cjc' provided by the NAG
library \cite{NAG}.
It is based on a variable order, variable step size Adam's method.  The
simulations either utilize periodic boundary conditions (to model droplets on
a ring-like track \cite{SMHY05})
or boundary conditions that mimick non-wettable borders of an otherwise
wettable channel (to model droplets on finite stripe-like tracks) \cite{SMHY05,Sumi05pre}. 
Simulations are started from steady droplet
solutions which develop in the absence of a chemical field, i.e.\ imposing
$\phi=0$. Then the droplet movement is initiated by breaking the symmetry of the
$\phi$-field by imposing a small gradient and starting the adsorption/desorption reaction.
After a short initial transient the running droplets follow periodic
trajectories that do not depend on details of the initial symmetry breaking.
For the periodic boundary condititions the initial solution was one stationary 
droplet, whereas in the case of non-wettable boundaries the initial solution
were two stationary droplets.

In the case of periodic boundary conditions the droplets move with constant
speed and shape after an initial phase. Fig.\,\ref{m2persim}\,(a) shows
space-time plots of the evolution of the film thickness for different reaction
rates $r$. The droplet velocity increases with the reaction rate as expected
from our continuation results.
Fig.\,\ref{m2persim}\,(b) shows a comparison of the droplet velocities
obtained in the simulations and by continuation. 
The values for the simulations are estimated after the
droplets have reached a constant speed and shape.
One finds that both velocities match fairly well. However, for higher reaction rates the
simulations slightly overestimate the velocities compared to the
continuation results. This results from the lower and equidistant
discretization used in the simulation in time.

\begin{figure}[hbt]
\begin{center}
\includegraphics[width=0.7\hsize]{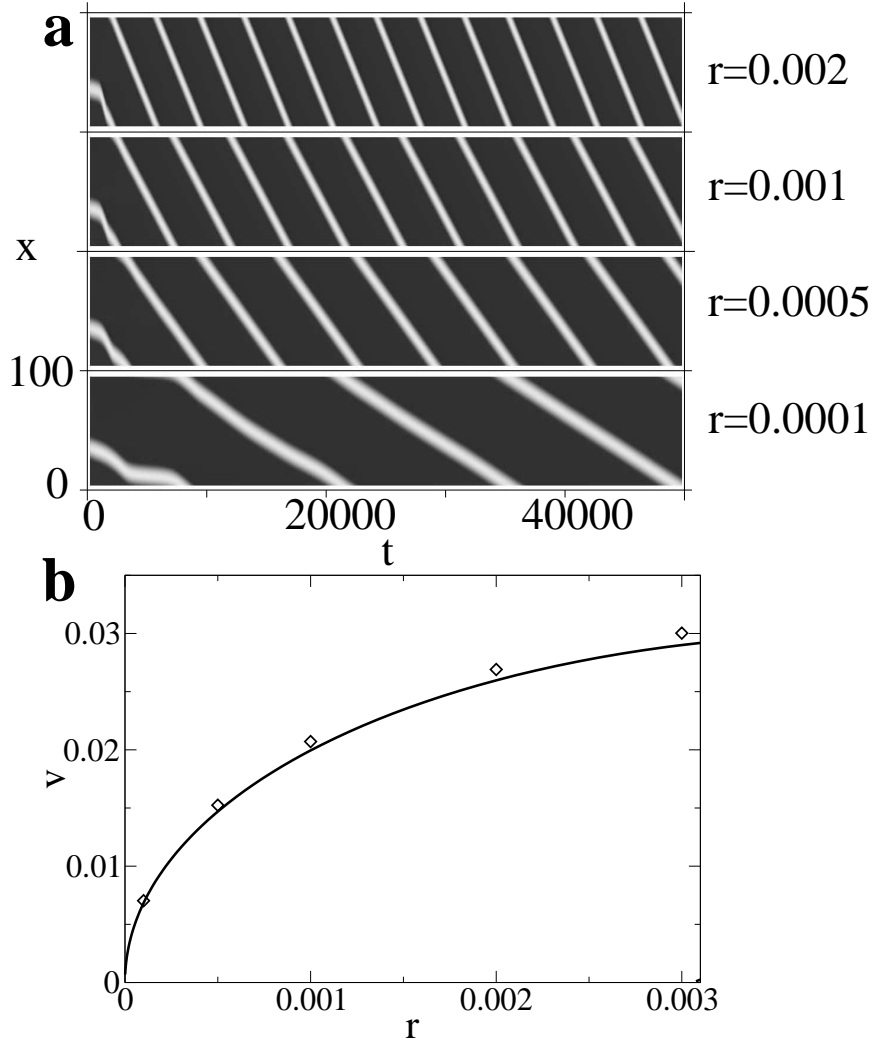}
\end{center}
\caption{Simulations of running droplets on a one-dimensional circular stripe (periodic boundary
conditions). Shown are (a) space time plots of the film thickness for different reaction
rates $r$ (indicated on the right) and (b) a comparison of drop velocities obtained by numerical
simulation ($\Diamond$) and continuation (solid line) for different values of the reaction rate $r$. Remaining
parameters are $g=2$, $s=1$, $d=0.001$, $L=100$, $\bar{h}=1.5$, $h_c=2$ and
$\Delta{}h=0.2$. Dark (light) colors correspond to small (large) film thickness.
\mylab{m2persim}
}
\end{figure}
\begin{figure}[hbt]
\begin{center}
\includegraphics[width=0.7\hsize]{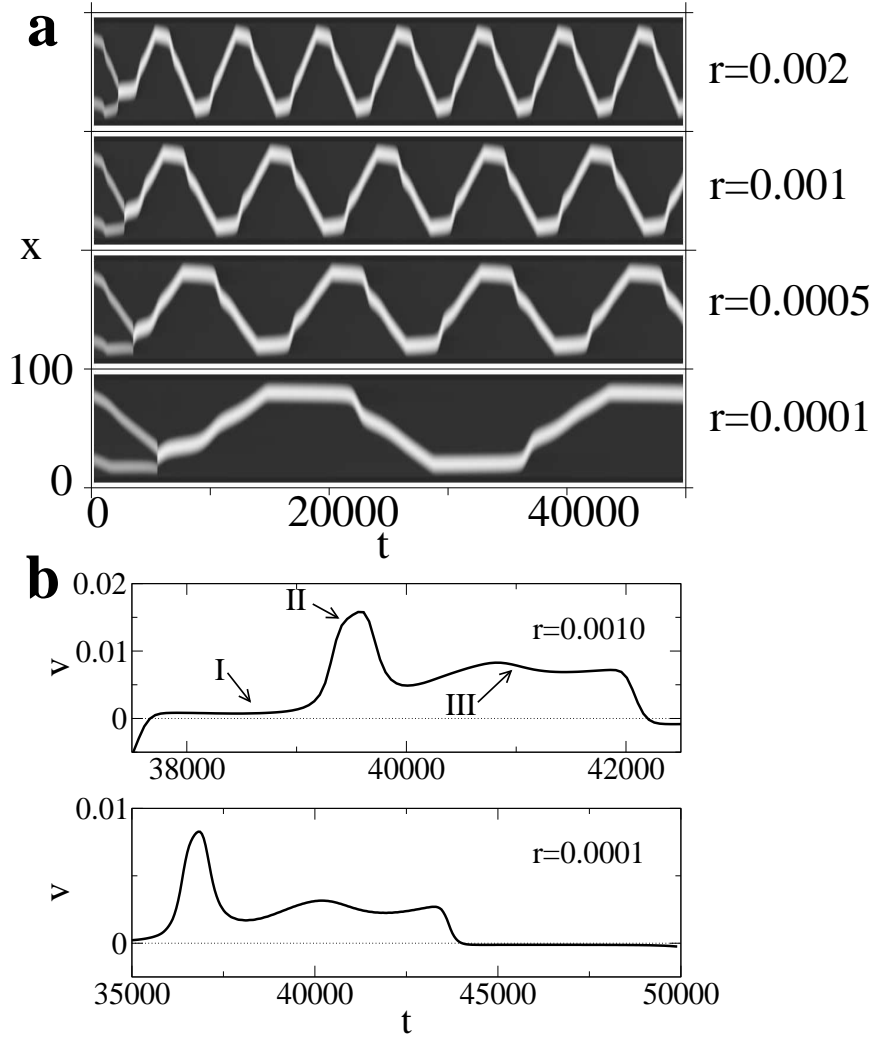}
\end{center}
\caption{Simulations of running droplets on a one-dimensional finite stripe bounded by
non-wettable borders. Shown are (a) space time plots of the film thickness for different reaction
rates $r$ (indicated on the right) and (b) the velocity of the center of mass of the film thickness depending
on time examplary for $r=0.001$ (top) and $r=0.0001$ (bottom). The thin dotted
line in (b) at $v=0$ is meant to guide the eye. Remaining
parameters are as in Fig.\,\ref{m2persim}. Dark (light) colors correspond to small (large) film thickness.
\mylab{m2refsim}
}
\end{figure}
\begin{figure}[hbt]
\begin{center}
\includegraphics[width=0.7\hsize]{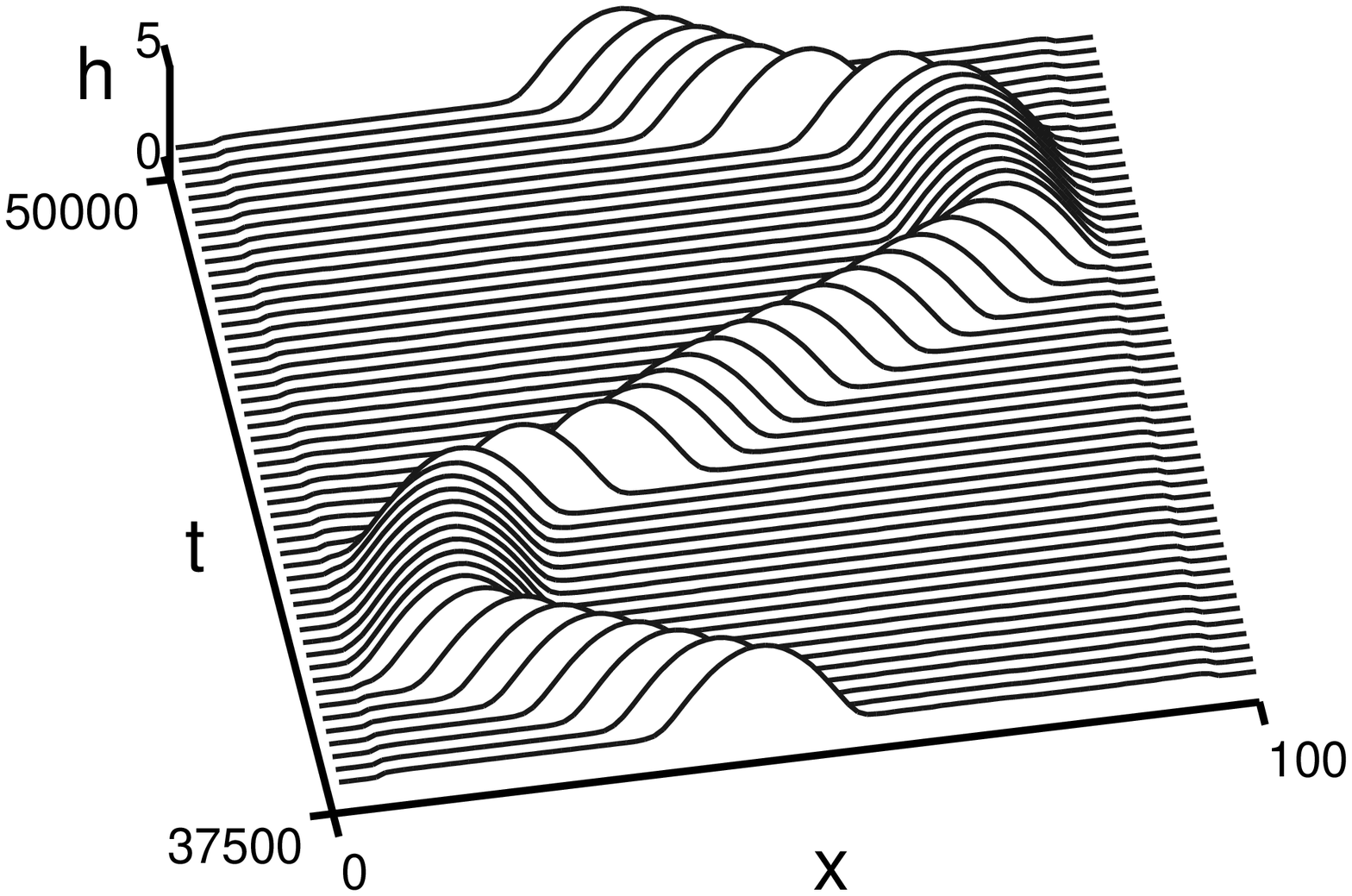}\\
\includegraphics[width=0.7\hsize]{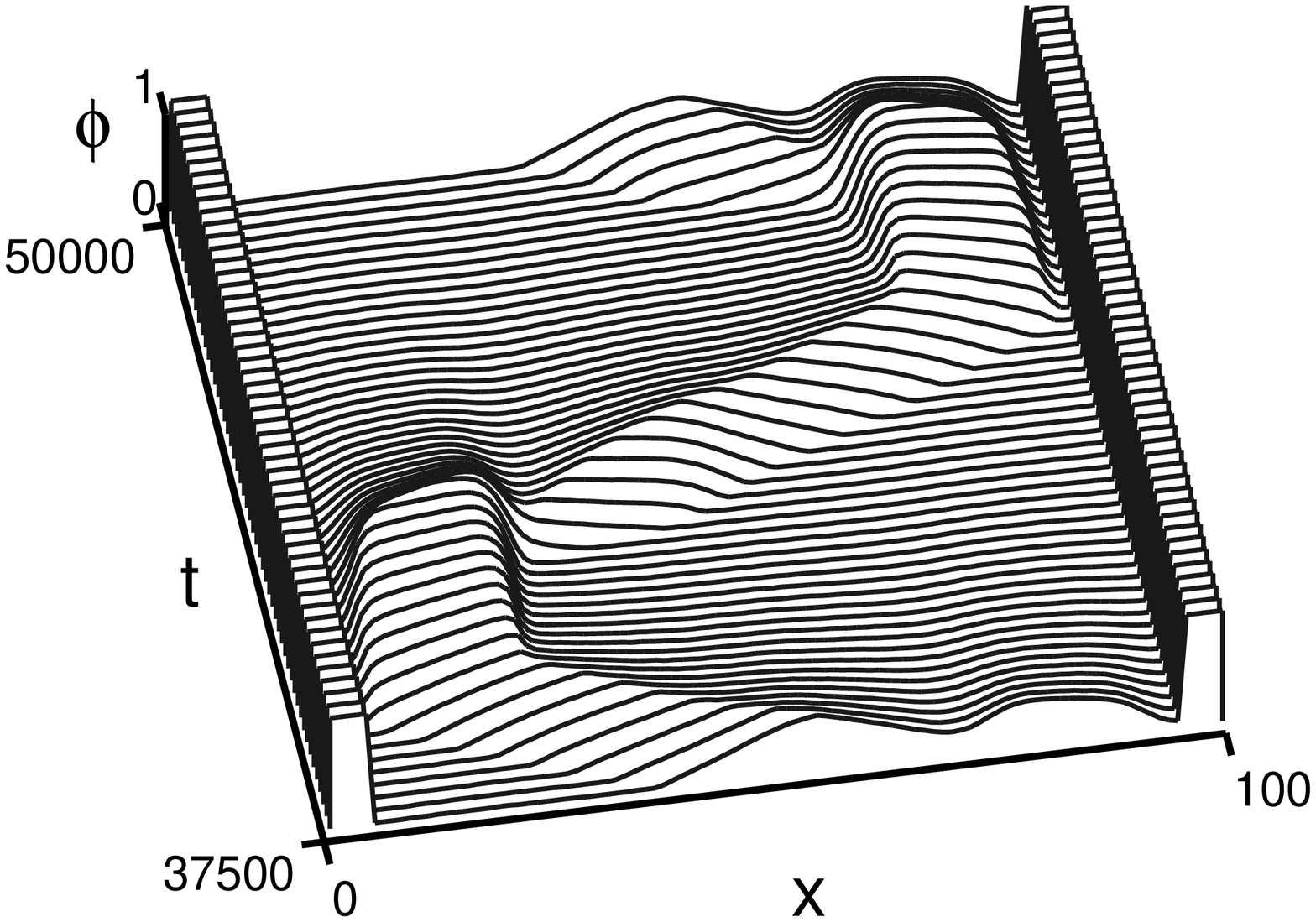}
\end{center}
\caption{Hidden-line space-time plots showing one halfcycle of the periodic movement
of a running droplet on a one-dimensional stripe bounded by
non-wettable borders. Shown are
the film thickness (top) and the $\phi$-field (bottom). $r=0.0005$ and remaining parameters are as in Fig.\,\ref{m2persim}.
\mylab{wf}
}
\end{figure}

Experiments with droplets in a finite wettable channel found 'regular rhythmic
motion' \cite{SMHY05} or different types of 'shuttling
motion' along with slowing and stopping behavior \cite{Sumi05pre}. 
The here performed simulations show a smooth transition between the different types
of periodic movements depending on our control parameters.

In general, the droplets in a wettable channel with non-wettable walls move periodically
between the two walls as shown for different reaction rates $r$ 
in the space-time plots in Fig.\,\ref{m2refsim}\,(a).
The initial solution of two stationary droplets quickly coarses upon starting
the chemical reactions. The prevailing droplet oscillates between the
channel walls with a frequency that depends on $r$.

The droplet movement can be classified into two identical but antisymmetric halfcycles
(i.e.\ with different signs in the velocity, and profiles that are related by
reflection). Fig.\,\ref{m2refsim}\,(b) shows the droplet velocity
depending on time during one halfcycle
exemplary for two reaction rates $r$. We find that
each half cycle typically contains three phases, distinguishable by their
different velocities.
After meeting the non-wettable boundary (phase I) the droplet velocity is very low or
even zero in the case of very small reaction rates $r$. In this phase, the $\phi$-field,
which has been produced by thof e passing droplet has first to decrease, until the
droplet can return on its own path. 
This phase of very small velocity is followed by a
short phase of a very high velocity (phase II) until the droplet returns to a medium
velocity (phase III), that is kept until it meets the opposite wall
and the next half cycle begins.
The subphases can also be very well distinguished in Fig.\,\ref{wf}, were we
show hidden line plots of the film thickness profiles (top) and coverage
profiles (bottom) for one period of droplet movement. One can
clearly see that in phase~I (velocity close to zero), after the droplet has
encountered the non-wettable wall, the concentration in the $\phi$-field is
very high and drops steeply in phase~II when the droplet starts moving again.    
Similar subphases of droplet motion have also been observed in 
Ref.\,\cite{Sumi05pre}. In the context of that work the continuous transition from 
Fig.\,\ref{m2refsim}\,(a) between $r=0.002$ (top) and $r=0.0001$ (bottom)
can be seen as their transition between 'shuttling motion' and 'intermittent
shuttling motion'.

A simple bead-spring model based on a mechanical analogy is used
in Ref.\,\cite{Sumi05pre}. It models the different experimentally
observed regimes varying the wettability of the walls. Although it well
captures the overall behaviour it can not resolve the more hydrodynamic aspects
of the motion like the flow field inside the running droplets or the 
dynamical contact angles.
%
\section{Conclusions}
\mylab{conc}
%
In the present work we have developed and analysed two models for chemically-driven 
self-propelled running droplets on solid substrates. Such moving droplets
were described for several experimental systems using different liquids, substrates
and reacting substances \cite{BBM94,DoOn95,LeLa00,LKL02,SMHY05,Sumi05pre}.
The movment of the droplets is driven by a self-produced wettability gradient 
that is perpetuated with the droplet itself by means of a desorption or adsorption reaction
underneath the droplet.

Two types of experimental systems were reported on: (I) adsorption underneath
the droplet decreases the wettability of the substrate irreversibly 
\cite{DoOn95,LeLa00,LKL02}; and (II) desorption underneath the droplet
removes a more wettable coating that is recovered behind the droplet 
through an adsorption from a surrounding medium \cite{SMHY05,Sumi05pre}.
We have described the droplet dynamics for both types of experimental systems using
coupled evolution equations for the profiles of the film thickness and the adsorbate coverage. The equations can be derived from the Navier-Stokes
equations using a long-wave or lubrication approximation \cite{ODB97} and
assuming a small Damk\"ohler number.
The two types of experiments have been mapped onto two types of models. In the 
type~I model a wettability-decreasing reaction takes place underneath the
droplet. In the type~II model this mechanism is extended by additionally introducing a
wettability-increasing reaction that takes place at the substrate outside the
droplet. 

The wettability of the substrate enters both models through a disjoining
pressure supplementing the Laplace pressure in the thin film equation. We
have chosen here a disjoining pressure consisting of a long-range
destabilizing part $\sim h^{-3}$ and a short-range stabilizing part 
$\sim h^{-6}$ used, for instance, in Ref.\,\cite{PiPo04} to study coarsening in
dewetting. The long-range part corresponds to van der Waals interaction
and is not influenced by the adsorbate. All the influence of the adsorbate
goes into the short-range part that in the simplest case selected here depends
linearly on the adsorbate coverage. 

%
%
Using continuation techniques and numerical simulations we have 
analyzed the solution behaviour of both models. Thereby we have 
focused on stationary running and sitting droplets 
in two dimensions. Both models display a transition 
from a non-saturated  to a saturated regime with increasing reaction rate. 
The transition is also obtained 
when increasing the droplet volume. In the non-saturated regime an increase 
in the reaction rate leads to a larger wettability 
gradient implying a larger droplet velocity.
In the saturated regime an increase in the reaction rate does not increase
the coverage at the rear of the droplet, i.e.\ it does neither lead to a larger
wettability gradient nor to  a larger droplet velocity.
However, it has turned out that the driving force and in consequence the
velocity are decreasing slightly with increasing reaction rate. This effect is due
to a rise in the adsorbat concentration in the advancing contact zone.
A similar behavior occurs when the droplet volume increases. There,
however, in the saturated regime the velocity clearly decreases with increasing
volume because the constant driving force (wettability gradient) is
counteracted by an increasing viscous friction. 
The latter dependencies found for the non-saturated and the saturated regime 
correspond very well to experimental results of 
Ref.\,\cite{DoOn95} (Fig.\,5) and Ref.\,\cite{LKL02} [Fig.\,7\,(a)], respectively. 
To our knowledge there exist, however, no experimental results for a
physico-chemical system that show the qualitative change between the 
two regimes in dependence of the droplet velocity on its volume. 
With the combination of materials used in Ref.\,\cite{LKL02} this should be
possible because their Fig.\,6\,(a) shows the transition from the non-saturated
to saturated regime for increasing solute concentration within the droplet.
This corresponds directly to the dependence on the reaction rate shown here in
Fig.\,\ref{m1vrcr}. 

Allowing for diffusion of the adsorbate along the substrate does not affect
the results in the non-saturated regime too much. However, it leads to a stronger
decrease of the velocity in the saturated regime because adsorbate is
transported to the substrate in front of the running droplet thereby decreasing the
overall wettability gradient. This may even lead to a transition towards sitting
droplets. This transition occurs either continuously through a supercritical drift-pitchfork
bifurcation or discontinuously through a saddle-node
bifurcation. In the latter case a metastable parameter region exist where
running and sitting drops can coexist. 

%
%
We have determined the existence regions of running droplets depending on the reaction
rate, the diffusion constant in the surface
coating, the droplet volume and the desorption/adsorption ratio (only for
model~II). Both models show a strong dependence of the existence regions of
sitting and running droplets on the diffusion constant and only a week
dependence on the droplet volume. 
The strong dependence on the diffusion constant results mainly 
from the diffusion of the $\phi$-field in front of the advancing droplet, thereby greatly
affecting the driving force and the droplet velocity.
The type~II model also shows a strong dependence on the desorption/adsorption
ratio, which is due to a decreased substrate coverage in close proximity behind the
droplet, effectively decreasing the driving force.  
The primary effect of the desorption reaction is the possibility for the
droplet to return on its own path, which we have illustrated by numerical simulations of
droplets moving on a wettable stripe with non-wettable borders in an
oscillatory manner, which has also been observed experimentally \cite{SMHY05,Sumi05pre}.
Finally, we have shown that different modes of periodic movement called 'shuttling motion' and 'intermittent
shuttling motion' in Ref.\,\cite{Sumi05pre} are covered by the presented model.

%
%
The analysis of both models has shown that for the chemically-driven running droplets
the advancing contact angle is always smaller than the receding
one. As was already shown in Ref.\,\cite{TJB04} also here the differences between
the static and the dynamic contact angles at the front and the rear 
are one order of magnitude smaller than the difference between
the two dynamic contact angles. This challenges the assumption of equal
dynamic contact angles at the front and the rear that was used in 
Ref.\,\cite{Brde95,deGe98}
to develop a simple description of self-propelled running droplets.
A simple quantitative theory should instead be based on the assumption that
the respective dynamic contact angles equal the different static contact
angles at the front and rear. 

%
%
%
The proposed type~I and type~II model based on thin film or lubrication theory
can very well reproduce the main features
of droplet motion that have been observed experimentally. However,
they fail to reproduce the reported damped oscillations in the
droplet shape overlaying the continuous droplet movement \cite{Sumi05pre}. 
These oscillations could be on the one hand the result of a weakly
inhomogeneous surface, since the authors 
of Ref.\,\cite{Sumi05pre} themselves suggested that they have no
full control of the experimental surface properties. On the other hand
the oscillations could be a sign of a Hopf bifurcation along the solution
branch of running droplets.
We are well aware that the models presented in this paper are minimal models
that, however, reproduce qualitatively most aspects of the behaviour of
chemically self-propelled droplets. Refinements of the presented theory
could include the viscous motion of the ambient medium as present in the 
type~II experiments. This extension can still be based on the lubrication
approximation, for instance, along the lines of the two-layer systems studied in 
Refs.\,\cite{MPBT05,PBMT04}. A second important extension could
cover the case of a higher Damk\"ohler number. Such a model has to
include a description of the transport of the solute within the droplet. 
%
%
%
\begin{appendix}
\section{Numerical techniques}
\subsection{Continuation}
\mylab{auto}
Sitting and running droplets are steady solutions in the laboratory and comoving
frame, respectively. They can be followed in parameter space using numerical 
continuation techniques \cite{DKK91,DKK91b}, for instance, 
employing the continuation software AUTO97 \cite{AUTO97}. 
The following section highlights some technical details of the
employed techniques.

The basic idea behind continuation is that unknown solutions of an algebraic
system for a certain set of control parameters are obtained by 
iterative techniques from known solutions nearby in parameter space.
Differential equations of the form
\begin{equation}
u'(x)=f(u(x),p)\qquad\mbox{with}\qquad f,u\,\in\,{\bf R}^n
\end{equation}
subject to initial, boundary and integral constraints
are discretized in space and then the resulting algebraic system is solved iterativly.
Here the dash indicates the first derivative with respect to $x$ and
$p$ denotes the set of control parameters.
The presence of boundary conditions and/or integral conditions requires the
presence of free parameters which are determined simultaneously and are part
of the solution to the differential equation.
The package AUTO97 is limited to the continuation of ordinary differential equations
(ODE's), thus it can only be used to compute droplet solutions in two dimensions.
As an example we consider the continuation of stationary running droplets, steady in
a comoving frame.
After transforming Eqs.\,(\ref{sys1}) and (\ref{sys2}) into the comoving frame
with velocity $v$
and integrating the resulting time-independent thin-film equation we have the
system of ODE's
\begin{eqnarray}
h'_1 & = & h_2 \mylab{as1}\\
h'_2 & = & h_3\\
h'_3 & = & \frac{\mu-vh_1}{h^3_1}-\Pi_x(h_1,\phi_1)\\
\phi'_1 & = & \phi_2\\
\phi'_2 & = & -\frac{1}{d}\left(R(h_1,\phi_1)+v\phi_2\right)\,,\mylab{as2}
\end{eqnarray}
%
%
where $\mu$ is an integration constant. It has the physical
meaning of a mean flow in the comoving frame.
$h_1$, $h_2$, $h_3$, $\phi_1$ and $\phi_2$ denote $h$,
$\partial_xh$, $\partial_{xx}h$, $\phi$ and
$\partial_x\phi$, respectively. The dashed quantities denote first derivatives with
respect to $x$.
The system is flux conservative, thus we need to specify the integral condition
\begin{equation}
0=\frac{1}{L}\int{}\,h_1\,dx-\bar{h}
\end{equation}
where $L$ is the system length and $\bar{h}$ is the mean film thickness.
Furthermore, for a complete description of the system we introduce periodic boundary conditions
for the film thickness
\begin{eqnarray}
h_1(0) & = & h_1(L)\\
h_2(0) & = & h_2(L)\\
h_3(0) & = & h_3(L)
\end{eqnarray}
and the mixed boundary conditions 
\begin{eqnarray}
0 & = & \xi(h_1(0))\lambda_1+\left\{\xi(h_1(0))+s\left[1-\xi(h_1(0))\right]\right\}\times\nonumber\\
    & & \left(\phi_2(0)-\phi_1(0)\lambda_1\right)\\
0 & = & \xi(h_1(0))\lambda_2+\left\{\xi(h_1(L))+s\left[1-\xi(h_1(L))\right]\right\}\times\nonumber\\
    & &\left(\phi_2(L)-\phi_1(L)\lambda_2\right)
\end{eqnarray}
for the $\phi$-field (see appendix \ref{boundary}). 
Since translation of a running droplet solution in the $x$-direction also
yields a valid solution one needs to introduce a pinning
condition in the form of an additional boundary or integral condition, which
will not be specified further in here.

As mentioned earlier, the number of boundary conditions (NBCD) and integral
conditions (NINT) imposes a constraint on the number of parameters (NPAR),
that have to be varied during the continuation process. 
Specifically NPAR=NBCD+NINT-NDIM+1, where NDIM is
the dimensionality of the system of ODE's.
This condition leaves us here with three free parameters, which are the
principal continuation parameter (e.g.\,the reaction rate $r$), the mean 
flow $\mu$ and the droplet velocity $v$.
AUTO97 uses the method of Orthogonal Collocation for discretizing
solutions, where the solution is approximated by piecewise polynomials with
2--7 collocation points per mesh interval. The mesh is adaptive as to
equidistribute the discretization error.
Having specified the ODE system in standard form with boundary and integral
conditions AUTO97 then tries to find stationary solutions to the discretized
system, by using a combination of Newton and Chord iterative methods.  
Once the solution has converged AUTO97 proceeds along the solution branch by a
small step in the parameter space defined by the free continuation parameters and restarts the iteration.\\
The challenge usually is, to provide AUTO97 with a nonuniform starting
solution for the continuation. 
For our purpose it is
sufficient to start the continuation close to the point of the primary 
bifurcation point, where the stable uniform solution of the ODE system 
(\ref{as1})--(\ref{as2}) or a system
similar to (\ref{as1})--(\ref{as2}) with periodic boundary 
conditions undergoes a Hopf-bifurcation. In the vicinity of the
bifurcation one can determine analytically small-amplitude sinusoidal stationary 
traveling waves.
By selectively using reaction, boundary or integral conditions as primary
continuation parameter one
finally computes fully nonlinear solutions for the film thickness and the coverage.\\
AUTO97 is not only able to follow solution branches but can also detect
bifurcations, like saddle-node bifurcation or branching points, and can then follow these
bifurcations in parameter space.
\subsection{Boundary conditions}
\mylab{boundary}
The use of periodic boundary conditions does not rule out interactions between
droplets in consecutive periods either via the film thickness or the
$\phi$-field. Interactions through the $\phi$-field arise if the
desorption of the coating is slow compared to the drop movement. One way to
avoid this problem is to use very large periods, such that the $\phi$-field
has enough time to recover. This approach is successful in suppressing
interactions but is very costly from a computational point of view. An
alternative approach is to use other than periodic boundary conditions for the
$\phi$-field. The following paragraphs illustrate this approach.

We generally consider a running droplet steady in a comoving frame, that has its
maximum approximately in the center of the computational domain of length $L$. 
We assume that the film thickness profile obeys periodic boundary conditions
$h_b\equiv h(0)=h(L)$ and $\partial_zh(0)=\partial_zh(L)\ll{}1$. The latter restriction ensures that 
the minimal film thickness is close to equilibrium and the period is large 
compared to the droplet.
\subsubsection{Model I}
\label{apm1}
In model~I we assume that there is no chemical reaction taking place outside the
droplet. Therefore, outside the computational domain for $x\le{}0$ and
$x\ge{}L$ the following ordinary differential equation holds
\begin{equation}
0=d\phi_{xx}+v\phi_x\,,\mylab{ode1}
\end{equation}
which has the general solution
\begin{equation}
\phi(x)=c+c'e^{-\frac{v}{d}x}
\end{equation}
with $c$ and $c'$ being yet undetermined constants.
For simplicity we assume that $v>0$, i.e.\,the droplet is moving to the
right.
In front of the droplet it is assumed that $\phi\rightarrow 0$ as
$x\rightarrow \infty$. Furthermore we assume that the $\phi$-field adopts a
finite value behind the droplet such that $\phi\rightarrow\phi_\infty$ as
$x\rightarrow{}-\infty$. 
First we consider the boundary at $x=0$ with the boundary value for the $\phi$-field
$\phi(0)=\phi_0$ and the first derivative of the $\phi$-field $\phi_x(0)=\phi_{x0}$.
Since $\phi_\infty$ is a finite value we find $c'=0$, which leads to the
Neumann boundary condition
\begin{equation}
\phi_{x0}=0\,.
\end{equation}
Next we consider the boundary at $x=L$ with the boundary value for the $\phi$-field
$\phi(L)=\phi_L$ and the first derivative of the $\phi$-field $\phi_x(L)=\phi_{xL}$.
Integrating (\ref{ode1}) we find 
\begin{equation}
0=d\phi_x+v\phi+c''\,,
\end{equation}
where $c''$ is an integration constant.
Since $\phi\rightarrow{}0$ as $x\rightarrow \infty$ and also
$\phi_x\rightarrow{}0$ as $x\rightarrow{}\infty$ we find $c''=0$, which leads
us to the mixed boundary condition
\begin{equation}
0=\phi_Lv+\phi_{xL}d\,.
\end{equation} 
\subsubsection{Model II}
\label{apm2}
We assume for the $h$-field outside the computational domain with $x\le{}0$ or $x\ge{}L$
$h(x)=h(0)=h(L)=h_b=const.$.
Then the concentration profile for the
$\phi$-field for $x\le{}0$ or $x\ge{}L$ can be computed analytically by solving a linear
2$^{nd}$ order ordinary differential equation in $x$ of the form
\begin{equation}
0=R_2(h_b,\phi)+d\phi_{xx}+v\phi_x\mylab{ode}\,,
\end{equation}
which has the general solution
\begin{equation}
\phi(x)=c+c'e^{\lambda_1{}x}+c''e^{\lambda_2{}x}\,,
\end{equation}
where
\begin{equation}
c=\frac{\xi(h_b)}{\xi(h_b)+s\left[1-\xi(h_b)\right]}
\end{equation}
\begin{equation}
\lambda_{1,2}=-\frac{v}{2d}\pm\sqrt{\frac{v^2}{4d^2}+\frac{r}{d}\left\{\xi(h_b)+s\left[1-\xi(h_b)\right]\right\}}
\end{equation}
and $c'$ and $c''$ are yet undetermined constants. For
$x\rightarrow{}\pm\infty$ $\phi$ adopts a small but finite value
$\phi_\infty$.

First we consider the boundary at $x=0$.  This leaves us with $c''=0$.
Using the boundary values for the $\phi$-field $\phi(0)=\phi_0$ and the first derivative of
the $\phi$-field $\phi_x(0)=\phi_{x0}$ at the boundary $x=0$ the following
system of equations holds
\begin{eqnarray}
0 & = & \phi_0-c-c'\\
0 & = & \phi_{x0}-c'\lambda_1\,.
\end{eqnarray}
Solving for $c'$ one finds the following mixed boundary condition
\begin{equation}
0=\xi(h_b)\lambda_1+\left\{\xi(h_b)+s\left[1-\xi(h_b)\right]\right\}\left(\phi_{x0}-\phi_0\lambda_1\right)\,.
\end{equation}
The same procedure can be performed at the boundary $x=L$ with
$\phi(L)=\phi_L$ and $\phi_x(L)=\phi_{xL}$ yielding the
condition
\begin{equation}
0=\xi(h_b)\lambda_2+\left\{\xi(h_b)+s\left[1-\xi(h_b)\right]\right\}\left(\phi_{xL}-\phi_L\lambda_2\right)
\end{equation}
In the continuation algorithm the two periodic boundary conditions for the
$\phi$-field were substituted by the two mixed boundary conditions obtained
above. 
\end{appendix}

\section*{Acknowledgments}

We thank
E. Knobloch for his comments on the drift-pitchfork bifurcation, and acknowledge support through the EU RTN ''Unifying principles in
non-equilibrium pattern formation'' (Contract MRTN-CT-2004--005728).


\begin{thebibliography}{10}
\providecommand*{\bibinfo}[2]{#2}
\providecommand*{\eprint}[1]{#1}
\providecommand*{\url}[1]{#1}
\bibitem{Newt1730hab2}
\bibinfo{author}{I.~Newton}, \bibinfo{title}{\emph{Opticks}}
  (\bibinfo{publisher}{G. Bell \& Sons LTD.}, London, \bibinfo{year}{1931}),
  (reprinted 4th ed.\ 1730), Book III, Part 1, Querie 31.
\bibitem{Hauk1710}
\bibinfo{author}{F.~Hauksbee}, \bibinfo{journal}{Phil. Trans.}
  \bibinfo{volume}{\textbf{27}}, \bibinfo{pages}{395} (\bibinfo{date}{1710}).
\bibitem{Vela98}
\bibinfo{author}{M.~G. Velarde}, \bibinfo{journal}{Philos. Trans. R. Soc. Lond.
  Ser. A-Math. Phys. Eng. Sci.} \bibinfo{volume}{\textbf{356}},
  \bibinfo{pages}{829} (\bibinfo{date}{1998}).
\bibitem{Broc89}
\bibinfo{author}{F.~Brochard}, \bibinfo{journal}{Langmuir}
  \bibinfo{volume}{\textbf{5}}, \bibinfo{pages}{432} (\bibinfo{date}{1989}).
\bibitem{Gree78}
\bibinfo{author}{H.~P. Greenspan}, \bibinfo{journal}{J. Fluid Mech.}
  \bibinfo{volume}{\textbf{84}}, \bibinfo{pages}{125} (\bibinfo{date}{1978}).
\bibitem{Raph88}
\bibinfo{author}{E.~Rapha{\"e}l}, \bibinfo{journal}{C. R. Acad. Sci. Ser. II}
  \bibinfo{volume}{\textbf{306}}, \bibinfo{pages}{751} (\bibinfo{date}{1988}).
\bibitem{ChWh92}
\bibinfo{author}{M.~K. Chaudhury} and \bibinfo{author}{G.~M. Whitesides},
  \bibinfo{journal}{Science} \bibinfo{volume}{\textbf{256}},
  \bibinfo{pages}{1539} (\bibinfo{date}{1992}).
\bibitem{ION00}
\bibinfo{author}{K.~Ichimura}, \bibinfo{author}{S.~K. Oh}, and
  \bibinfo{author}{M.~Nakagawa}, \bibinfo{journal}{Science}
  \bibinfo{volume}{\textbf{288}}, \bibinfo{pages}{1624} (\bibinfo{date}{2000}).
\bibitem{JJP03}
\bibinfo{author}{J.~F. Joanny}, \bibinfo{author}{F.~J{\"u}licher}, and
  \bibinfo{author}{J.~Prost}, \bibinfo{journal}{Phys. Rev. Lett.}
  \bibinfo{volume}{\textbf{90}}, \bibinfo{pages}{168102}
  (\bibinfo{date}{2003}).
\bibitem{Vent1799}
\bibinfo{author}{Venturi}, \bibinfo{journal}{Ann. de Chimie}
  \bibinfo{volume}{\textbf{XXI}}, \bibinfo{pages}{262} (\bibinfo{date}{1799}).
\bibitem{Toml1869}
\bibinfo{author}{C.~Tomlinson}, \bibinfo{journal}{Phil. Mag. Ser. 4}
  \bibinfo{volume}{\textbf{46}}, \bibinfo{pages}{409} (\bibinfo{date}{1869}).
\bibitem{Rayl1890b}
\bibinfo{author}{L.~Rayleigh}, \bibinfo{journal}{Proc. R. Soc. London}
  \bibinfo{volume}{\textbf{47}}, \bibinfo{pages}{364} (\bibinfo{date}{1890}).
\bibitem{Haya02}
\bibinfo{author}{Y.~Hayashima}, \bibinfo{author}{M.~Nagayama},
  \bibinfo{author}{Y.~Doi}, \bibinfo{author}{S.~Nakata},
  \bibinfo{author}{M.~Kimura}, and \bibinfo{author}{M.~Iida},
  \bibinfo{journal}{Phys. Chem. Chem. Phys.} \bibinfo{volume}{\textbf{4}},
  \bibinfo{pages}{1386} (\bibinfo{date}{2002}).
\bibitem{CMS64}
\bibinfo{author}{R.~L.~Cottington}, \bibinfo{author}{C.~M.~Murphy} 
and \bibinfo{author}{C.~R.~Singleterry},
  \bibinfo{journal}{Adv. Chem. Ser.} \bibinfo{volume}{\textbf{43}},
  \bibinfo{pages}{341} (\bibinfo{date}{1964}).
\bibitem{BiQu00}
\bibinfo{author}{J.~Bico} and \bibinfo{author}{D.~Qu\'er\'e},
  \bibinfo{journal}{Europhys. Lett.} \bibinfo{volume}{\textbf{51}},
  \bibinfo{pages}{546} (\bibinfo{date}{2000}).
\bibitem{Mara1871}{C.~G.~Marangoni}, \bibinfo{journal}{Ann. Phys. (Poggendorf)}
  \bibinfo{volume}{\textbf{143}}, \bibinfo{pages}{337} (\bibinfo{date}{1871}),
Observation\,16.
\bibitem{RRV94}
\bibinfo{author}{A.~Y. Rednikov}, \bibinfo{author}{Y.~S. Ryazantsev}, and
  \bibinfo{author}{M.~G. Velarde}, \bibinfo{journal}{Phys. Fluids}
  \bibinfo{volume}{\textbf{6}}, \bibinfo{pages}{451} (\bibinfo{date}{1994}).
\bibitem{Rieg03}
\bibinfo{author}{H.~Riegler}, \bibinfo{journal}{personal communication}
  (\bibinfo{date}{2003}).
\bibitem{YoPi05_pre}
\bibinfo{author}{A.~Yochelis} and \bibinfo{author}{L.~M. Pismen},
  \bibinfo{journal}{Phys. Rev. E} \bibinfo{volume}{\textbf{72}}, \bibinfo{pages}{025301(R)} (\bibinfo{date}{2005}).
\bibitem{BBM94}
\bibinfo{author}{C.~D. Bain}, \bibinfo{author}{G.~D. Burnetthall}, and
  \bibinfo{author}{R.~R. Montgomerie}, \bibinfo{journal}{Nature}
  \bibinfo{volume}{\textbf{372}}, \bibinfo{pages}{414} (\bibinfo{date}{1994}).
\bibitem{DoOn95}
\bibinfo{author}{F.~Domingues Dos~Santos} and
  \bibinfo{author}{T.~Ondar{\c{c}}uhu}, \bibinfo{journal}{Phys. Rev. Lett.}
  \bibinfo{volume}{\textbf{75}}, \bibinfo{pages}{2972} (\bibinfo{date}{1995}).
\bibitem{LeLa00}
\bibinfo{author}{S.~W. Lee} and \bibinfo{author}{P.~E. Laibinis},
  \bibinfo{journal}{J. Am. Chem. Soc.} \bibinfo{volume}{\textbf{122}},
  \bibinfo{pages}{5395} (\bibinfo{date}{2000}).
\bibitem{LKL02}
\bibinfo{author}{S.~W. Lee}, \bibinfo{author}{D.~Y. Kwok}, and
  \bibinfo{author}{P.~E. Laibinis}, \bibinfo{journal}{Phys. Rev. E}
  \bibinfo{volume}{\textbf{65}}, \bibinfo{pages}{051602}
  (\bibinfo{date}{2002}).
\bibitem{SMHY05}
\bibinfo{author}{Y.~Sumino}, \bibinfo{author}{N.~Magome},
  \bibinfo{author}{T.~Hamada}, and \bibinfo{author}{K.~Yoshikawa},
  \bibinfo{journal}{Phys. Rev. Lett.} \bibinfo{volume}{\textbf{94}}(6),
  \bibinfo{pages}{068301}, \bibinfo{eid}{068301} (\bibinfo{numpages}{4}  pages)
   (\bibinfo{date}{2005}).
\bibitem{Sumi05pre}
\bibinfo{author}{Y.~Sumino}, \bibinfo{author}{H.~Kitahata},
  \bibinfo{author}{K.~Yoshikawa}, \bibinfo{author}{M.~Nagayama},
  \bibinfo{author}{S.~M. Nomura}, \bibinfo{author}{N.~Magome}, and
  \bibinfo{author}{Y.~Mori}, \bibinfo{journal}{Phys. Rev. E} \bibinfo{volume}{\textbf{72}},
  \bibinfo{pages}{041603} (\bibinfo{date}{2005}).
\bibitem{Mage03}
\bibinfo{author}{R.~Magerle}, \bibinfo{journal}{personal communication}
  (\bibinfo{date}{2003}).
\bibitem{SBH00}
\bibinfo{author}{A.~K. Schmid}, \bibinfo{author}{N.~C. Bartelt}, and
  \bibinfo{author}{R.~Q. Hwang}, \bibinfo{journal}{Science}
  \bibinfo{volume}{\textbf{290}}, \bibinfo{pages}{1561} (\bibinfo{date}{2000}).
\bibitem{LaEu96}
\bibinfo{author}{K.~Landry} and \bibinfo{author}{N.~Eustathopoulos},
  \bibinfo{journal}{Acta Mater.} \bibinfo{volume}{\textbf{44}},
  \bibinfo{pages}{3923} (\bibinfo{date}{1996}).
\bibitem{Yost98}
\bibinfo{author}{F.~G. Yost}, \bibinfo{journal}{Scr. Mater.}
  \bibinfo{volume}{\textbf{38}}, \bibinfo{pages}{1225} (\bibinfo{date}{1998}).
\bibitem{WBR98}
\bibinfo{author}{J.~A. Warren}, \bibinfo{author}{W.~J. Boettinger}, and
  \bibinfo{author}{A.~R. Roosen}, \bibinfo{journal}{Acta Mater.}
  \bibinfo{volume}{\textbf{46}}, \bibinfo{pages}{3247} (\bibinfo{date}{1998}).
\bibitem{VMHE99}
\bibinfo{author}{R.~Voitovitch}, \bibinfo{author}{A.~Mortensen},
  \bibinfo{author}{F.~Hodaj}, and \bibinfo{author}{N.~Eustathopoulos},
  \bibinfo{journal}{Acta Mater.} \bibinfo{volume}{\textbf{47}},
  \bibinfo{pages}{1117} (\bibinfo{date}{1999}).
\bibitem{SCT00}
\bibinfo{author}{E.~Saiz}, \bibinfo{author}{R.~M. Cannon}, and
  \bibinfo{author}{A.~P. Tomsia}, \bibinfo{journal}{Acta Materialia}
  \bibinfo{volume}{\textbf{48}}, \bibinfo{pages}{4449} (\bibinfo{date}{2000}).
\bibitem{WeGr02}
\bibinfo{author}{W.~B. Webb} and \bibinfo{author}{G.~S. Grest},
  \bibinfo{journal}{Scr. Mater.} \bibinfo{volume}{\textbf{47}},
  \bibinfo{pages}{393} (\bibinfo{date}{2002}).
\bibitem{ZWT98}
\bibinfo{author}{D.~W. Zheng}, \bibinfo{author}{W.~Wen}, and
  \bibinfo{author}{K.~N. Tu}, \bibinfo{journal}{Phys. Rev. E}
  \bibinfo{volume}{\textbf{57}}, \bibinfo{pages}{R3719} (\bibinfo{date}{1998}).
\bibitem{KRE99}
\bibinfo{author}{S.~Kalogeropoulou}, \bibinfo{author}{C.~Rado}, and
  \bibinfo{author}{N.~Eustathopoulos}, \bibinfo{journal}{Scr. Mater.}
  \bibinfo{volume}{\textbf{41}}, \bibinfo{pages}{723} (\bibinfo{date}{1999}).
\bibitem{Brde95}
\bibinfo{author}{F.~Brochard-Wyart} and \bibinfo{author}{P.-G. de~Gennes},
  \bibinfo{journal}{C. R. Acad. Sci. Ser. II} \bibinfo{volume}{\textbf{321}},
  \bibinfo{pages}{285} (\bibinfo{date}{1995}).
\bibitem{deGe98}
\bibinfo{author}{P.-G. de~Gennes}, \bibinfo{journal}{Physica A}
  \bibinfo{volume}{\textbf{249}}, \bibinfo{pages}{196} (\bibinfo{date}{1998}).
\bibitem{MiMe97}
\bibinfo{author}{A.~Mikhailov} and \bibinfo{author}{D.~Meink{\"o}hn}, in
  \emph{Lecture Notes in Physics} (\bibinfo{publisher}{Springer},
  \bibinfo{year}{1997}), \bibinfo{volume}{vol. 484}, \bibinfo{pages}{pp.
  334--345}.
\bibitem{deGe99}
\bibinfo{author}{P.-G. de~Gennes}, \bibinfo{journal}{Comptes Rendus de
  l'Academie des Sciences, Serie II} \bibinfo{volume}{\textbf{327}},
  \bibinfo{pages}{147} (\bibinfo{date}{1999}).
\bibitem{deGe97}
\bibinfo{author}{P.~G. de~Gennes}, \bibinfo{journal}{Europhys. Lett.}
  \bibinfo{volume}{\textbf{39}}, \bibinfo{pages}{407} (\bibinfo{date}{1997}).
\bibitem{TJB04}
\bibinfo{author}{U.~Thiele}, \bibinfo{author}{K.~John}, and
  \bibinfo{author}{M.~B{\"a}r}, \bibinfo{journal}{Phys. Rev. Lett.}
  \bibinfo{volume}{\textbf{93}}, \bibinfo{pages}{027802}
  (\bibinfo{date}{2004}).
\bibitem{ODB97}
\bibinfo{author}{A.~Oron}, \bibinfo{author}{S.~H. Davis}, and
  \bibinfo{author}{S.~G. Bankoff}, \bibinfo{journal}{Rev. Mod. Phys.}
  \bibinfo{volume}{\textbf{69}}, \bibinfo{pages}{931} (\bibinfo{date}{1997}).
\bibitem{deGe85}
\bibinfo{author}{P.-G. de~Gennes}, \bibinfo{journal}{Rev. Mod. Phys.}
  \bibinfo{volume}{\textbf{57}}, \bibinfo{pages}{827} (\bibinfo{date}{1985}).
\bibitem{Hunt92}
\bibinfo{author}{R.~J. Hunter}, \bibinfo{title}{\emph{Foundation of Colloid
  Science}}, \bibinfo{volume}{vol.~1} (\bibinfo{publisher}{Clarendon Press},
  Oxford, \bibinfo{year}{1992}).
\bibitem{Isra92}
\bibinfo{author}{J.~N. Israelachvili}, \bibinfo{title}{\emph{Intermolecular and
  Surface Forces}} (\bibinfo{publisher}{Academic Press}, London,
  \bibinfo{year}{1992}).
\bibitem{TNPV02}
\bibinfo{author}{U.~Thiele}, \bibinfo{author}{K.~Neuffer},
  \bibinfo{author}{Y.~Pomeau}, and \bibinfo{author}{M.~G. Velarde},
  \bibinfo{journal}{Colloid Surf. A} \bibinfo{volume}{\textbf{206}},
  \bibinfo{pages}{135} (\bibinfo{date}{2002}).
\bibitem{Prob94}
\bibinfo{author}{R.~F. Probstein}, \bibinfo{title}{\emph{Physicochemical
  Hydrodynamics}} (\bibinfo{publisher}{Wiley}, New York, \bibinfo{year}{1994}),
  2nd ed.
\bibitem{disjpress}
Note, that the disjoining pressure used in Ref.\,\cite{TJB04} was
  $\Pi(h)=\frac{2 S_a d_0^2}{h^3} +
  \frac{S_p}{l}\,\left(1+\frac{\phi}{g}\right)\exp\left[\frac{d_0-h}{l}\right]$
  where for $\phi=0$, $S_a$ and $S_p$ are the apolar and polar components of
  the total spreading coefficient $S=S_a+S_p$, respectively, and $l$ is a
  correlation length \cite{Shar93}. One usually describes the choice $S_a>0$ and $S_p<0$ as a combination of a stabilizing long-range van der Waals and a
  destabilizing short-range polar interaction. The apparent contradiction of
  qualitative similar results for model~I for different verbal descriptions and
  combinations of signs used here and in Ref.\,\cite{TJB04} results from
  a subtle feature of the combination of exponential and power law. Combining a
  term $\sim 1/h^3$ and one $\sim \exp (-h)$ leads for a proper choice of
  parameters to a dominance of $1/h^3$ for large and very small $h$. The
  exponential only dominates for intermediate thicknesses (see
  Ref.\,\cite{TVN01} for a related phase diagram). This implies that
  above verbal description only covers part of the feature of the disjoining
  pressure. On the contrary, the combination of two power laws used here
  clearly attributes the long range and short range forces to the terms
  $h^{-3}$ and $h^{-6}$, respectively. We therefore believe, that the chosen
  disjoining pressure more accurately represents the physical situation.
\bibitem{Shar93}
\bibinfo{author}{A.~Sharma}, \bibinfo{journal}{Langmuir}
  \bibinfo{volume}{\textbf{9}}, \bibinfo{pages}{861} (\bibinfo{date}{1993}).
\bibitem{DKK91}
\bibinfo{author}{E.~Doedel}, \bibinfo{author}{H.~B. Keller}, and
  \bibinfo{author}{J.~P. Kernevez}, \bibinfo{journal}{Int. J. Bif. Chaos}
  \bibinfo{volume}{\textbf{1}}, \bibinfo{pages}{493} (\bibinfo{date}{1991}).
\bibitem{DKK91b}
\bibinfo{author}{E.~Doedel}, \bibinfo{author}{H.~B. Keller}, and
  \bibinfo{author}{J.~P. Kernevez}, \bibinfo{journal}{Int. J. Bif. Chaos}
  \bibinfo{volume}{\textbf{1}}, \bibinfo{pages}{745} (\bibinfo{date}{1991}).
\bibitem{AUTO97}
\bibinfo{author}{E.~J. Doedel}, \bibinfo{author}{A.~R. Champneys},
  \bibinfo{author}{T.~F. Fairgrieve}, \bibinfo{author}{Y.~A. Kuznetsov},
  \bibinfo{author}{B.~Sandstede}, and \bibinfo{author}{X.~J. Wang},
  \bibinfo{title}{\emph{AUTO97: Continuation and bifurcation software for
  ordinary differential equations}} (\bibinfo{publisher}{Concordia University},
  Montreal, \bibinfo{year}{1997}).
\bibitem{Thie02}
\bibinfo{author}{U.~Thiele}, \bibinfo{author}{K.~Neuffer},
  \bibinfo{author}{M.~Bestehorn}, \bibinfo{author}{Y.~Pomeau}, and
  \bibinfo{author}{M.~G. Velarde}, \bibinfo{journal}{Colloid Surf. A -
  Physicochem. Eng. Asp.} \bibinfo{volume}{\textbf{206}}, \bibinfo{pages}{87}
  (\bibinfo{date}{2002}).
\bibitem{KTB92}
\bibinfo{author}{M.~Kness}, \bibinfo{author}{L.~S. Tuckerman}, and
  \bibinfo{author}{D.~Barkley}, \bibinfo{journal}{Phys. Rev. A}
  \bibinfo{volume}{\textbf{46}}, \bibinfo{pages}{5054} (\bibinfo{date}{1992}).
\bibitem{KrMi94}
\bibinfo{author}{K.~Krischer} and \bibinfo{author}{A.~Mikhailov},
  \bibinfo{journal}{Phys. Rev. Lett.} \bibinfo{volume}{\textbf{73}},
  \bibinfo{pages}{3165} (\bibinfo{date}{1994}).
\bibitem{HaMe94}
\bibinfo{author}{A.~Hagberg} and \bibinfo{author}{E.~Meron},
  \bibinfo{journal}{Chaos} \bibinfo{volume}{\textbf{4}}, \bibinfo{pages}{477}
  (\bibinfo{date}{Sept. 1994}).
\bibitem{OBSP98}
\bibinfo{author}{M.~Or-Guil}, \bibinfo{author}{M.~Bode}, \bibinfo{author}{C.~P.
  Schenk}, and \bibinfo{author}{H.-G. Purwins}, \bibinfo{journal}{Phys. Rev. E}
  \bibinfo{volume}{\textbf{57}}, \bibinfo{pages}{6432} (\bibinfo{date}{1998}).
\bibitem{Boed03}
\bibinfo{author}{H.~U. B\"odeker}, \bibinfo{author}{M.~C. R\"ottger},
  \bibinfo{author}{A.~W. Liehr}, \bibinfo{author}{T.~D. Frank},
  \bibinfo{author}{R.~Friedrich}, and \bibinfo{author}{H.-G. Purwins},
  \bibinfo{journal}{Phys. Rev. E} \bibinfo{volume}{\textbf{67}},
  \bibinfo{pages}{056220} (\bibinfo{date}{2003}).
\bibitem{PrJo88}
\bibinfo{author}{M.~R.~E. Proctor} and \bibinfo{author}{C.~A. Jones},
  \bibinfo{journal}{J. Fluid Mech.} \bibinfo{volume}{\textbf{188}},
  \bibinfo{pages}{301} (\bibinfo{date}{1988}).
\bibitem{CGG89}
\bibinfo{author}{P.~Coullet}, \bibinfo{author}{R.~E. Goldstein}, and
  \bibinfo{author}{G.~H. Gunaratne}, \bibinfo{journal}{Phys. Rev. Lett.}
  \bibinfo{volume}{\textbf{63}}, \bibinfo{pages}{1954} (\bibinfo{date}{1989}).
\bibitem{KnMo90}
\bibinfo{author}{E.~Knobloch} and \bibinfo{author}{D.~R.~Moore},
  \bibinfo{journal}{Phys. Rev. A} \bibinfo{volume}{\textbf{42}},
  \bibinfo{pages}{4693} (\bibinfo{date}{1990}).
\bibitem{RiPa92}
\bibinfo{author}{H.~Riecke} and \bibinfo{author}{H.~G. Paap},
  \bibinfo{journal}{Phys. Rev. A} \bibinfo{volume}{\textbf{45}},
  \bibinfo{pages}{8605} (\bibinfo{date}{1992}).
\bibitem{ThKn04}
\bibinfo{author}{U.~Thiele} and \bibinfo{author}{E.~Knobloch},
  \bibinfo{journal}{Physica D} \bibinfo{volume}{\textbf{190}},
  \bibinfo{pages}{213} (\bibinfo{date}{2004}).
\bibitem{ThKn03}
\bibinfo{author}{U.~Thiele} and \bibinfo{author}{E.~Knobloch},
  \bibinfo{journal}{Phys. Fluids} \bibinfo{volume}{\textbf{15}},
  \bibinfo{pages}{892} (\bibinfo{date}{2003}).
\bibitem{MPBT05}
\bibinfo{author}{D.~Merkt}, \bibinfo{author}{A.~Pototsky},
  \bibinfo{author}{M.~Bestehorn}, and \bibinfo{author}{U.~Thiele},
  \bibinfo{journal}{Phys. Fluids} \bibinfo{volume}{\textbf{17}},
  \bibinfo{pages}{064104} (\bibinfo{date}{2005}).
\bibitem{NAG}
\bibinfo{title}{\emph{{NAG C} library, {M}ark 6}} (\bibinfo{date}{2000}),
  www.nag.co.uk.
\bibitem{PiPo04}
\bibinfo{author}{L.~M. Pismen} and \bibinfo{author}{Y.~Pomeau},
  \bibinfo{journal}{Phys. Fluids} \bibinfo{volume}{\textbf{16}},
  \bibinfo{pages}{2604} (\bibinfo{date}{2004}).
\bibitem{PBMT04}
\bibinfo{author}{A.~Pototsky}, \bibinfo{author}{M.~Bestehorn},
  \bibinfo{author}{D.~Merkt}, and \bibinfo{author}{U.~Thiele},
  \bibinfo{journal}{Phys. Rev. E} \bibinfo{volume}{\textbf{70}},
  \bibinfo{pages}{025201(R)} (\bibinfo{date}{2004}).
\bibitem{TVN01}
\bibinfo{author}{U.~Thiele}, \bibinfo{author}{M.~G. Velarde}, and
  \bibinfo{author}{K.~Neuffer}, \bibinfo{journal}{Phys. Rev. Lett.}
  \bibinfo{volume}{\textbf{87}}, \bibinfo{pages}{016104}
  (\bibinfo{date}{2001}).

\end{thebibliography}
\end{document}